\documentclass{article}
\usepackage{amsmath,amsfonts}

\usepackage{stfloats}
\usepackage{amsmath,amssymb,amsfonts}
\usepackage{graphicx}
\usepackage{textcomp}
\usepackage[table]{xcolor}
\def\BibTeX{{\rm B\kern-.05em{\sc i\kern-.025em b}\kern-.08em
T\kern-.1667em\lower.7ex\hbox{E}\kern-.125emX}}

\usepackage{hyperref}
\hypersetup{
colorlinks = true, 
linkcolor=black, 
citecolor=black, 
urlcolor=black 
}

\usepackage{paralist}
\usepackage{subcaption}
\usepackage{amsmath}
\usepackage{amssymb}
\usepackage{pifont}
\usepackage[most]{tcolorbox}
\usepackage{booktabs}
\usepackage{xspace}
\usepackage{bbding}
\usepackage{cite}
\usepackage{multirow}
\usepackage{multicol}
\usepackage{amsthm}
\theoremstyle{definition}

\providecommand{\definitionname}{Definition}
\theoremstyle{definition}
\newtheorem{example}{\protect\examplename}
\providecommand{\examplename}{Example}

\newcommand{\terminateExample}{\hfill $\lhd$}

\newcommand{\method}{\textit{BootQA}\xspace}
\newcommand{\transpose}{\mathrm{T}}
\newcommand{\originalTestSuite}{\ensuremath{\mathit{TS}}\xspace}
\newcommand{\testCase}[1]{\ensuremath{t_{#1}}\xspace}
\newcommand{\availTestCaseNum}{\ensuremath{s}\xspace}
\newcommand{\propNum}{\ensuremath{r}\xspace}
\newcommand{\propSet}{\ensuremath{P}\xspace}
\newcommand{\prop}[1]{\ensuremath{p_{#1}}\xspace}

\newcommand{\propValue}[2]{\ensuremath{v_{#1}^{#2}}\xspace} 
\newcommand{\theoLimit}[1]{\ensuremath{L}^{#1}\xspace}
\newcommand{\weight}[1]{\ensuremath{w}_{#1}\xspace}
\newcommand{\testCaseVector}{\ensuremath{\boldsymbol{t}}\xspace}
\newcommand{\xVector}{\ensuremath{\boldsymbol{x}}\xspace}
\newcommand{\propValueVector}[1]{\boldsymbol{v}^{#1}}
\newcommand{\et}{\ensuremath{\mathtt{et}}\xspace}
\newcommand{\fr}{\ensuremath{\mathtt{fr}}\xspace}
\newcommand{\num}{\ensuremath{\mathtt{num}}\xspace}
\newcommand{\gs}{\textit{GSDTSR}\xspace}
\newcommand{\io}{\textit{IOF/ROL}\xspace}
\newcommand{\pa}{\textit{PaintControl}\xspace}
\newcommand{\subsetDataset}[1]{\ensuremath{Subset_{#1}}\xspace}

\newcommand{\fv}{\ensuremath{\mathit{FV}}\xspace}
\newcommand{\nq}{\ensuremath{\mathit{NQ}}\xspace}
\newcommand{\objFunc}{\ensuremath{\mathcal{O}}}
\newcommand{\objFuncOverall}{\ensuremath{\mathcal{O}_{\mathit{all}}}}
\newcommand{\numOfSampledSubsets}{\ensuremath{M}\xspace}
\newcommand{\subproblemSize}{\ensuremath{N}\xspace}
\newcommand{\ext}{\textit{exTime}\xspace}

\usepackage{algpseudocode}
\usepackage{algorithm}

\tcbuselibrary{breakable}
\newtcbtheorem{Summary}{Conclusion}{
breakable,
theorem name,
enhanced,
coltitle=black,
top=0.1in,
left=0in,
attach boxed title to top left=
{xshift=1.5em,yshift=-\tcboxedtitleheight/2},
boxed title style={size=small,colback=gray!60}
}{summary}

\usepackage{PRIMEarxiv}

\title{Test Case Minimization with Quantum Annealers}

\author{
  Xinyi~Wang\\
  Simula Research Laboratory \\
  Oslo, Norway\\
  \texttt{xinyi@simula.no} \\
   \And
  Asmar~Muqeet\\
  Simula Research Laboratory \\
  University of Oslo \\
  Oslo, Norway\\
  \texttt{asmar@simula.no} \\
   \And
  Tao~Yue \\
  Simula Research Laboratory \\
  Oslo, Norway\\
  \texttt{tao@simula.no} \\
   \And
  Shaukat~Ali \\
  Simula Research Laboratory and \\
  Oslo Metropolitan University \\
  Oslo, Norway\\
  \texttt{shaukat@simula.no} \\
   \And
  Paolo~Arcaini \\
  National Institute of Informatics \\
  Tokyo, Japan\\
  \texttt{arcaini@nii.ac.jp} \\
}

\begin{document}
\maketitle

\begin{abstract}
Quantum annealers are specialized quantum computers for solving combinatorial optimization problems using special characteristics of quantum computing (QC), such as superposition, entanglement, and quantum tunneling. Theoretically, quantum annealers can outperform classical computers. However, the currently available quantum annealers are small-scale, i.e., they have limited quantum bits (qubits); hence, they currently cannot demonstrate the quantum advantage. Nonetheless, research is warranted to develop novel mechanisms to formulate combinatorial optimization problems for quantum annealing (QA). However, solving combinatorial problems with QA in software engineering remains unexplored. Toward this end, we propose \method, the very first effort at solving the test case minimization (TCM) problem with QA. In \method, we provide a novel formulation of TCM for QA, followed by devising a mechanism to incorporate bootstrap sampling to QA to optimize the use of qubits. We also implemented our TCM formulation in three other optimization processes: classical simulated annealing (SA), QA without problem decomposition, and QA with an existing D-Wave problem decomposition strategy, and conducted an empirical evaluation with three real-world TCM datasets. Results show that \method outperforms QA without problem decomposition and QA with the existing decomposition strategy in terms of effectiveness. Moreover, \method's effectiveness is similar to SA. Finally, \method has higher efficiency in terms of time when solving large TCM problems than the other three optimization processes.
\end{abstract}


\section{Introduction}\label{introduction}

Quantum Computing (QC) is pledging to unravel complex computational problems more efficiently than classical computing~\cite{speed}. One particular application of QC in the near future is solving combinatorial optimization problems. To this end, quantum annealers -- a special kind of quantum computers have been used to demonstrate solving combinatorial optimization problems with quantum annealing (QA)~\cite{siloi2021investigating, perdomo2015quantum, inoue2020model}. To this end, D-Wave Systems Inc.\footnote{https://www.dwavesys.com/} has commercialized quantum annealers and continuously increases the number of quantum bits (qubits). Currently, the most powerful system from D-Wave is the \textit{Advantage} system with more than 5,000 qubits.
Theoretically, QA can solve optimization problems faster than state-of-the-art algorithms on classical computers~\cite{farhi2000quantum}. However, due to the small scale of currently available quantum annealers, demonstrating their quantum advantage over the classical is practically impossible. Current research challenges include devising novel problem encoding mechanisms and optimally utilizing the limited number of available qubits. With this purpose, some research has already experimentally demonstrated the potential of using QA for solving combinatorial optimization problems~\cite{siloi2021investigating, perdomo2015quantum, inoue2020model}. However, QA has not been investigated for solving optimization problems in software engineering, including test optimization.
Test optimization aims to improve the cost-effectiveness of software testing. Particularly, in the context of regression testing, test optimization can be applied to select and prioritize a subset of test cases for testing a new software version from the test suite used for testing previous software versions. Test optimization can involve various optimization objectives, such as minimizing the number of test cases to select to improve the test efficiency and maximizing the fault detection rate of the selected test cases to enhance the test effectiveness. In the literature, a range of test optimization solutions have been proposed~\cite{Survey1,survey2}, and search algorithms (e.g., NSGA-II~\cite{nsga}) have been commonly used for cost-effectively solving various test optimization problems~\cite{ARRIETA2019137,panichella2014improving,wang2015cost}.

In this paper, we aim to address the test case minimization (TCM) problem with QA, which aims to find a minimum number of test cases for execution~\cite{survey2}. However, two main research challenges exist. First, considering that QA solves an optimization problem by formulating it as an \textit{Ising Model} or a \textit{Quadratic Unconstrained Binary Optimization (QUBO)} model, we, hence, need a novel formulation for TCM. Second, current quantum annealers have a limited number of qubits. Therefore, they cannot be directly employed to solve complex TCM problems with large numbers of test cases, and new methods are needed to optimize the use of qubits.

In this paper, we propose \method to address the above two challenges. In particular, first, we propose a novel and generic QUBO formulation for TCM problems. Second, to optimize the use of qubits to solve large TCM problems, we introduce \textit{bootstrap sampling}~\cite{bootstrap} to the QA process. To investigate the cost-effectiveness of \method, we also implemented our QUBO formulation in three other optimization processes: classical Simulated Annealing (SA), QA without sub-problem decomposing, and QA with an existing decomposing strategy of D-Wave. We perform an empirical evaluation on three real-world datasets and evaluate the four optimization processes regarding their effectiveness and time efficiency in solving the three TCM problems.
Results show that \method exhibits similar effectiveness with \textit{SA} and outperforms the other two Quantum Processing Unit (QPU) based processes. Moreover, \method shows the highest time efficiency among all, especially for large-scale TCM problems. We provide the experiment results and implementations in the online repository: \url{https://github.com/AsmarMuqeet/BootQA}.

In the rest of the paper, Sect.~\ref{sec:background} introduces the background, and Sect.~\ref{sec:relatedwork} discusses the related work. Sect.~\ref{sec:method} presents the proposed test case minimization approach with QA. Sect.~\ref{sec:experiment_design} presents the empirical evaluation, and Sect.~\ref{sec:experimentResults} presents the evaluation results, followed by discussions in Sect.~\ref{discussions}. We conclude the paper in Sect.~\ref{sec:conclusions}.

\section{Background}\label{sec:background}
QA is based on the adiabatic theorem~\cite{adiabatic_theorem,QA_theory}, stating that if a quantum system is initialized in a \emph{ground state} (i.e., the lowest energy state) and the \emph{Hamiltonian}\footnote{In quantum mechanics, the Hamiltonian of a quantum system specifies its total energy~\cite{farhi2000quantum}, i.e., the sum of both the kinetic and potential energy of the system.} of the system changes sufficiently slowly, the quantum system will remain in the ground state during the system evolution. The adiabatic theorem can solve optimization problems by preparing a quantum system in a ground state of an easy-to-implement initial Hamiltonian and slowly evolving the initial Hamiltonian to the final Hamiltonian, whose ground state encodes solutions that solve the optimization problem. Below, we introduce the three main steps of solving an optimization problem with QA on a D-Wave QPU as described in~\cite{Yarkoni_2022}.

\noindent\textbf{Step 1: Objective function and QUBO formalization.}
A minimization objective function is defined as a mathematical expression representing the energy of the quantum system, which is converted into an Ising~\cite{kadowaki1998quantum} or QUBO model to be implemented on QA hardware. For D-Wave QA, QUBO is the standard model, which is a quadratic formulation with binary variables and formatted as a real-valued upper-diagonal weight matrix $Q$ with its dimensions defined by the number of binary variables~\cite{QUBO}, defined as follows:
%
\begin{equation}\label{eq:qubo}
\min f(\xVector) =\xVector^{\transpose} Q \xVector= \sum_iQ_{i,i}x_i + \sum_{i<j}Q_{i,j}x_ix_j
\end{equation}
where \xVector is the vector of binary decision variables, and $\xVector^{\transpose}$ is the transpose of \xVector. Diagonal terms $Q_{i, i}$ in $Q$ are the linear coefficients, and off-diagonal terms $Q_{i,j}$ are quadratic coefficients. For example, for a three-variable $(x_0, x_1, x_2)$ problem, the generic QUBO formulation is as follows:
\begin{equation*}
\min f(\xVector)= \begin{bmatrix}
x_0\,x_1\,x_2
\end{bmatrix} 
\begin{bmatrix}
Q_{00} & Q_{01} & Q_{02}\\
0 & Q_{11} & Q_{12}\\
0 & 0 & Q_{22}
\end{bmatrix} 
\begin{bmatrix}
x_0\\
x_1\\
x_2
\end{bmatrix} 
\end{equation*}

\begin{figure}[!tb]
\begin{subfigure}[t]{0.47\columnwidth}
\centering
\includegraphics[width=0.7\textwidth]{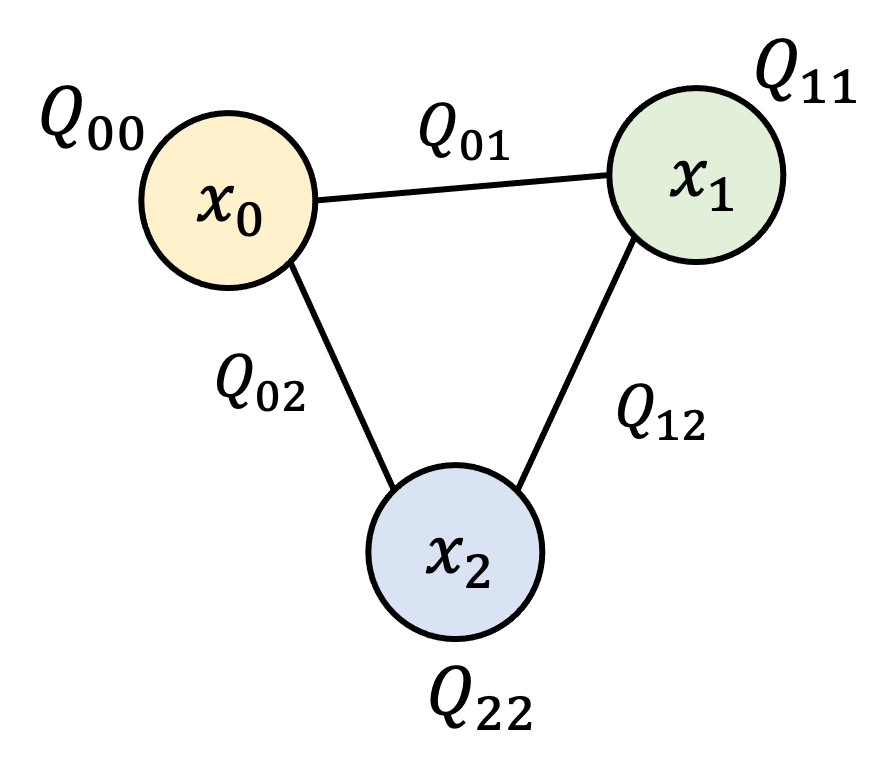}
\caption{QUBO graph}
\label{fig:QUBO_quboGraph}
\end{subfigure}
\hfill
\begin{subfigure}[t]{0.49\columnwidth}
\centering
\includegraphics[width=0.75\textwidth]{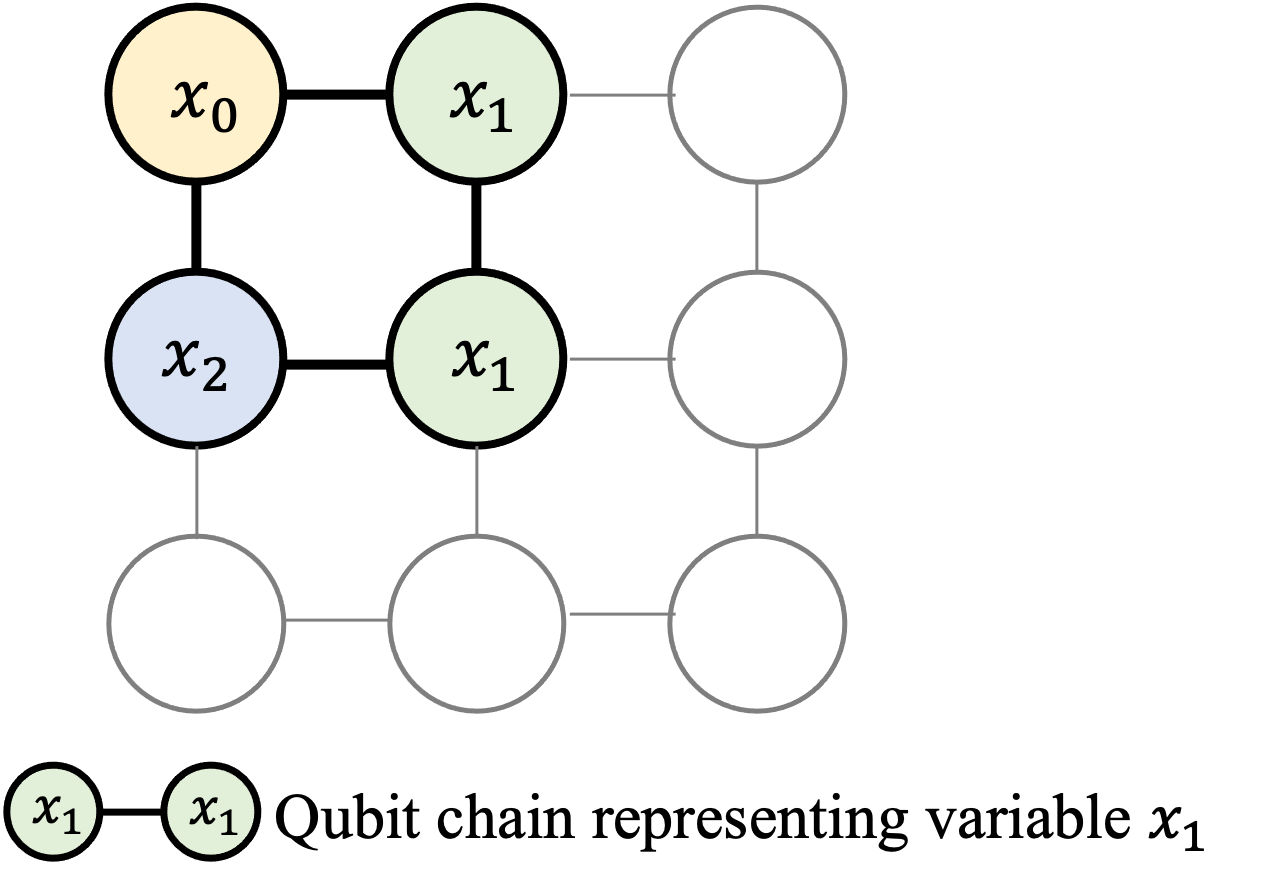}
\caption{Working graph}
\label{fig:QUBO_workingGraph}
\end{subfigure}
\caption{QUBO formalization for a three-variable problem}
\label{fig:QUBO}
\end{figure}

With Eq.~\ref{eq:qubo}, we convert an optimization problem into a \emph{QUBO graph} where each variable is represented as a node, and each non-zero off-diagonal term is represented as an edge between a pair of nodes connecting the corresponding variables, as shown, for example, in Fig.~\ref{fig:QUBO_quboGraph}.

\noindent\textbf{Step 2: Minor-embedding.}
We map a QUBO graph into a \textit{working graph} describing how physical qubits are organized and connected in the QA hardware. The process is called \textit{minor-embedding}, e.g., \textit{Chimera topology} for the \textit{D-Wave 2000Q} system and \textit{Pegasus} for the \textit{Advantage} system~\cite{techqpu,techqpu2020}. In a working graph (e.g., Fig.~\ref{fig:QUBO_workingGraph}), each node represents a \textit{physical qubit}, and each edge connecting a pair of nodes (i.e., physical qubits) denotes a \textit{coupler}. Ideally, each node in the QUBO graph (e.g., Fig.~\ref{fig:QUBO_quboGraph}) should be mapped to a physical qubit, and each edge should be mapped to a coupler. However, since the \textit{working graph} might not be fully connected, not all QUBO problems can be perfectly embedded into QA. For instance, to directly map the full-connected QUBO graph in Fig.~\ref{fig:QUBO_quboGraph} onto the working graph in Fig.~\ref{fig:QUBO_workingGraph}, direct physical connections between any two pairs of three qubits representing three variables are needed, which is not the case for this working graph. To handle this, D-Wave introduced \textit{chain}, which groups multiple physically connected qubits that can represent a node together (a variable in the QUBO formulation) of the QUBO graph. All the physical qubits in one chain act together as a single logical qubit to solve the optimization problem. In this example, $x_1$ from the QUBO graph (Fig.~\ref{fig:QUBO_quboGraph}) is mapped to two physical qubits in the working graph, i.e., a chain. As a result, each pair of variables is now connected in the working graph using this chain.

Minor embedding is time-consuming and requires specific algorithms to find suitable solutions~\cite{Choi2008}. A significant quantity of qubits and couplers is needed to embed some large and densely connected QUBO graphs. In this paper, we used the algorithms implemented by D-Wave for minor embedding.

\noindent\textbf{Step 3: Programming, initialization, and sampling.}
In the programming process, coefficients of the defined QUBO model are involved in defining the final Hamiltonian of the quantum system. Each linear coefficient is applied as \textit{weight qubit bias}, while each quadratic coefficient is mapped as \textit{coupler strength} in the QA hardware. After programming, all qubits are initialized in equal superposition as the initial Hamiltonian of the system.

QA is a heuristic algorithm. As a result, it cannot guarantee finding the optimal solution, i.e., reaching the system's ground state. Thus, multiple sampling processes are required to generate multiple candidate solutions in an execution. The number of sampling processes is determined by the \textit{number of reads} parameter in D-Wave QPU-API. Each sampling process has an annealing process and a read-out process. In the annealing process, the quantum system evolves from the initial to the final Hamiltonian in a time-dependent manner according to the Schr{\"o}dinger equation. When the system reaches the final Hamiltonian's ground state, all qubits are read out, and the values represent a candidate solution. Generally, we consider the solution with the lowest energy as the optimal solution generated by QA.

\section{Related Work}\label{sec:relatedwork}

\textit{Quantum annealing for optimization.} 
Several QA applications in various domains have emerged recently, such as portfolio problem~\cite{rosenberg2016solving}, machine learning~\cite{kulkarni2021quantum, nath2021review, hibat2021variational}, routing problems (e.g., traveling salesman~\cite{warren2013adapting, warren2020solving}, vehicle routing~\cite{SYRICHAS201752}, control of traffic signal~\cite{inoue2021traffic, hussain2020optimal}, process job scheduling~\cite{denkena2021quantum, stollenwerk2021agile}), and logistical network design~\cite{ding2021implementation}. Though QA has shown promising results in solving optimization problems in these domains, the potential for test case optimization remains unexplored, which is the focus of this paper.

\textit{Test case optimization for classical software.}
Test case optimization is often classified into test case prioritization and TCM~\cite{surveyclass}. Test case prioritization aims to identify the most optimal ordering of test cases for execution, while TCM revolves around selecting a minimum subset of test cases that minimizes the overall test execution cost while maximizing the overall fault detection capability, etc. Test case optimization has been extensively studied in the literature~\cite{tcs1, tcs2, tcs3}. For instance, the authors of the survey in~\cite{Survey1} identified and studied 90 test case optimization approaches and concluded that search-based (evolutionary algorithms, search algorithms) and clustering techniques are the most commonly used techniques.
Another survey~\cite{survey2} analyzes 29 TCM and prioritization approaches with machine learning applied. The survey concludes that these approaches mainly employ supervised learning, reinforcement learning, and natural language processing-based methods. Current machine learning-based approaches in software development pipelines with continuous integration are unsuitable since they require the reconstruction of machine learning models from scratch from time to time.
In contrast to existing works in the literature, we focus on using quantum annealers for TCM. Thus, our work is the first work on using quantum computers for solving TCM problems. 

\textit{Quantum software testing.}
In quantum software engineering~\cite{zhao2020quantum,Serrano2022bookQSE}, quantum software testing~\cite{MiranskyyICSE2020,MiranskyyICSE19,Garcia2023} is an active research sub-area. Several quantum software testing approaches have been proposed, such as based on search algorithms~\cite{search,mutation2,fuzz}, mutation testing~\cite{mutation,mutation1,Rui_mutation,Fortunato2022}, fuzzing~\cite{fuzz}, metamorphic testing~\cite{Rui_metamorphic,morphq}, and combinatorial testing~\cite{Combinatorial}, which all focus on generating test cases for testing quantum programs. In contrast, in this paper, we propose \method to minimize test suites for testing classical software with quantum computers. 

\section{Quantum annealing for test case minimization}\label{sec:method}
%
\begin{figure*}[!tb]
\centering
\includegraphics[width=0.9\textwidth]{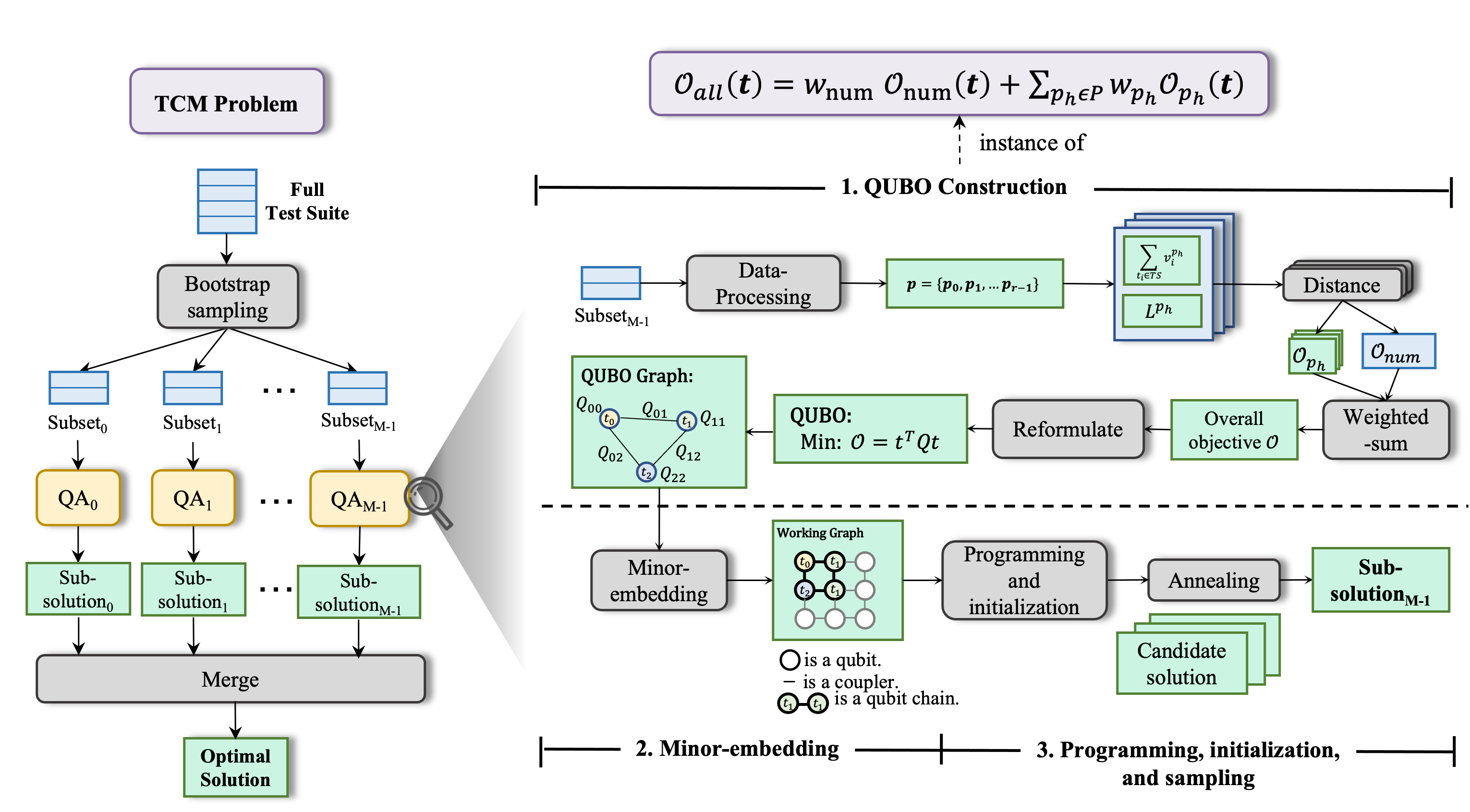}
\caption{Overview of \method}
\label{fig:overview}
\end{figure*}

This section, first, introduces the overview of the \method (Sect.~\ref{sec:bootstrap}). 
Second, we define the generic QUBO formulation (Sect.~\ref{sec:objective}) and illustrate it with the running example.

\subsection{Overview of \method}\label{sec:bootstrap}
A QUBO formulation can be represented as a QUBO graph (see Sect.~\ref{sec:background}). If a QUBO graph has many nodes and connections among them, a substantial amount of qubits and couplers in the hardware are needed to embed the whole problem, which is not be supported by today's QA hardware. To circumvent this issue, we use a bootstrap sampling strategy (Fig.~\ref{fig:overview}, left side) to decompose the original TCM problem into smaller sub-problems. Generally, bootstrap sampling creates small sub-problems by sampling subsets of test cases from the entire test suite to ensure that each sub-problem is solvable on the QA hardware. Note that each subset is created from the entire test suite; hence each test case may appear in different subsets, and all subsets are independent.

Specifically, given a test suite \originalTestSuite, we decompose the original test suite into \numOfSampledSubsets smaller test suites of size \subproblemSize as shown in Figure \ref{fig:overview}. \subproblemSize is given by considering the maximum number of qubits and couplers supported by a given QA hardware. We randomly sample \numOfSampledSubsets subsets of \originalTestSuite of size \subproblemSize, i.e., \subsetDataset{0}, \subsetDataset{1}, \ldots, \subsetDataset{M-1}, repeatedly until a certain percentage of \originalTestSuite is covered. This coverage percentage is a hyper-parameter $\beta$ of \method. These subsets are used as the input to formulate the sub-problems according to the QUBO formulation (Sect.~\ref{sec:objective}). Each sub-problem is solved individually with QA, and as a result, a corresponding sub-solution is produced. 

For each sub-problem (Fig.~\ref{fig:overview}, right side), we construct an overall objective function based on test cases and their properties in the sub-problem, which is an instance of the generic QUBO formulation introduced in Sect.~\ref{sec:objective}. We first extract the properties of test cases in each subset as optimization objectives. Second, we combine them into the overall objective function as a QUBO formulation (Step 1). 
With the QUBO graph converted from the formulation, we embed it into the hardware working graph (Step 2: minor-embedding). After the process of programming, initialization and sampling (Step 3), among all candidate solutions, we choose the solution with the lowest energy as the sub-solution of the sub-problem (see details in Sect.~\ref{sec:background}).

Eventually, for \numOfSampledSubsets sub-problems, we obtain \numOfSampledSubsets sub-solutions, each of which is a subset of the original test suite. Each selected test case in each sub-problem is considered as a selected test case in the final solution for the whole problem. In other words, we merge all the unique test cases selected in all sub-solutions to form the final solution for the whole problem. 


\subsection{QUBO Formulation for Test Case Minimization}\label{sec:objective}
TCM selects the smallest possible subset of test cases out of the total set of test cases available, for the software under test (SUT), while satisfying all testing objectives as much as possible. Examples of objectives include minimizing the overall test execution time and maximizing the fault detection rate of the selected subset of test cases, whose corresponding values can be obtained based on the history executions. For TCM, the selection of each test case is represented as a binary variable \testCase{i}. If \testCase{i} takes the value of 1, it means that the $i$-th test case is selected and 0 otherwise. Consequently, we represent the vector of $s$ variables in the QUBO as $\testCaseVector=[\testCase{0},$ $\testCase{1},$ $\ldots,$ $\testCase{s-1}]$.

Each test case is characterized with \propNum properties. Each property is associated with one testing objective. We represent the set of \propNum properties as $\propSet = \{\prop{0},$ $\prop{1},$ $\ldots,$ $\prop{r-1}\}$.

\begin{example}
To explain our approach, we use a small-scale running example composed of two test cases, shown in Table~\ref{table:running_example}.
\begin{table}[!tb]
\centering
\small
\caption{Running Example}
\label{table:running_example}
\renewcommand\arraystretch{1.2}
\resizebox{0.65\columnwidth}{!}{
\begin{tabular}{c|c|c}
\hline
Test Case ID & Execution Time \et (s) & Failure Rate \fr (\%)\\
\hline
\testCase{0} & 10 & 80\%\\
\testCase{1} & 20 & 40\%\\
\hline
\end{tabular}
}
\end{table}
Each test case has two properties: \textit{execution time} (\et) and \textit{failure rate} (\fr), shown as two columns in Table~\ref{table:running_example}. The values of these two properties are based on the history of test executions. The test case execution time is the average time a test case takes across all its executions. The failure rate is the percentage of times a test case failed out of the total number of executions during the history executions. For example, test case \testCase{0} has \textit{execution time} of 10 seconds and \textit{failure rate} of 80\%.
\terminateExample
\end{example}

For the $h$-th property, the vector of the values of all test cases is represented as $\propValueVector{\prop{h}}=$ $[\propValue{0}{\prop{h}},$ $\propValue{1}{\prop{h}},$ $\ldots,$ $\propValue{s-1}{\prop{h}}]$, where \propValue{i}{\prop{h}} represents the $h$-th property value of \testCase{i}. A TCM approach maximizes the overall effectiveness while minimizing the overall cost with the selected test cases. To this end, some properties shall be maximized (e.g., \textit{failure rate} in our running example), and others shall be minimized (e.g., \textit{execution time} in our running example). To treat all objectives equally in the overall objective function, we do the min-max normalization for properties whose ranges are not between 0 and 1.

For a property \prop{h}, we aim to optimize the cumulative sum of property values for the selected test cases. Specifically, we maximize the sum of property values for all the selected test cases if \prop{h} is an \emph{effectiveness property}, and conversely, we minimize the sum if \prop{h} is \emph{cost property}.
To do this, we define the objective function $\objFunc_{\prop{h}}$ (to be minimized) that computes the distance between the sum of values of the property for all selected test cases and the corresponding property's theoretical limit $\theoLimit{\prop{h}}$ with the Euclidean distance; $\theoLimit{\prop{h}}$ is the upper limit for an effectiveness property, and the lower limit for a cost property. Formally:
\begin{equation*}
\objFunc_{\prop{h}}(\testCaseVector)={(\propValueVector{\prop{h}}}\cdot{\testCaseVector}^{\transpose}-\theoLimit{\prop{h}})^2
\end{equation*}
where the sum of values of the property is calculated with matrix multiplication $\propValueVector{\prop{h}}\cdot{\testCaseVector}^{\transpose}$, with ${\testCaseVector}^{\transpose}$ being the transpose of $\testCaseVector$.
For an effectiveness property \prop{h}, we calculate the theoretical limit as the sum of values of all the test cases for \prop{h}. For a cost property, one possible theoretical limit is 0. However, depending on the context, other limits could be used. Since for each variable $\testCase{i}\in\{0, 1\}$, it holds that $\testCase{i}^2=\testCase{i}$, the above formula can then be expanded as below:
\begin{equation*}
\objFunc_{\prop{h}}(\testCaseVector) =
\sum_{i=0}^{\availTestCaseNum-1}({\propValue{i}{\prop{h}}}^2 - 2\propValue{\prop{i}{h}}{\prop{h}})\testCase{i}
+ 2 
\sum^{\availTestCaseNum-1}_{i<j}\propValue{i}{\prop{h}}\propValue{j}{\prop{h}}\cdot\testCase{i}\testCase{j}
+
{(\theoLimit{\prop{h}})}^2
\end{equation*}

\begin{example}
For the objective of \textit{failure rate} (i.e., the property \fr) of our running example, we represent the vector of values as $\propValueVector{\fr}=[\propValue{0}{\fr}, \propValue{1}{\fr}]$. The vector of two test cases is represented as $\testCaseVector=[\testCase{0}, \testCase{1}]$. The upper theoretical limit is $\theoLimit{\fr}$.
The objective formulation turns out to be:
%
\begin{equation*}
\begin{array}{ll} 
\objFunc_{\fr}(\testCaseVector) & =(\propValueVector{\fr}\cdot{\testCaseVector}^{\transpose}-\theoLimit{\fr})^2=(\propValue{0}{\fr}\testCase{0}+\propValue{1}{\fr}\testCase{1}-\theoLimit{\fr})^2 \\
& = ({\propValue{0}{\fr}}^2 - 2\theoLimit{\fr}\propValue{0}{\fr})\testCase{0} +({\propValue{1}{\fr}}^2 - 2\theoLimit{\fr}\propValue{1}{\fr})\testCase{1}\\
 & + 2\propValue{0}{\fr}\propValue{1}{\fr}\testCase{0}\testCase{1}+({\theoLimit{\fr}})^2
\end{array}
\end{equation*}
Since $\propValueVector{\fr}=$ $[0.8,$ $0.4]$ and $\theoLimit{\fr}=1.2$, the formula above is instantiated as:
\begin{equation*}
\objFunc_{\fr}(\testCaseVector)= -1.28\testCase{0}-0.8\testCase{1}+0.64\testCase{0}\testCase{1}+1.44
\end{equation*}
Similarly, for the objective of \textit{\et}, we first do normalization for the property values. Then we get the vector $\propValueVector{\et}=[\propValue{0}{\et}, \propValue{1}{\et}]=[0.5, 1]$ and $\theoLimit{\et}=0$. The objective function becomes:
\begin{equation*}
\objFunc_{\et}(\testCaseVector) = 0.25\testCase{0} + \testCase{1} + \testCase{0}\testCase{1}
\end{equation*}
\terminateExample
\end{example}

A key objective of TCM is to minimize the number of test cases selected (we name it as \num), i.e., the size of the selected test suite should be as small as possible. We add up the variable values in \testCaseVector ($\testCase{i}\in\{0,1\}$ and $\testCase{i}=1$ denoting that \testCase{i} is selected) to calculate \num. To be consistent with the formalization of the other two properties, we represent the total number of test cases selected as a matrix multiplication: $\propValueVector{\num}\cdot{\testCaseVector}^{\transpose}$, where all values in $\propValueVector{\num}$ are 1.
%
Since we want to minimize the number of selected test cases, we consider its lower theoretical limit $\theoLimit{\num}=0$. Consequently, the objective is represented as follows:
\begin{equation*}
\objFunc_{\num}(\testCaseVector) =(\propValueVector{\num}\cdot\testCaseVector^{\transpose}-\theoLimit{\num})^2
=\sum_{i=0}^{\availTestCaseNum-1}\testCase{i}+2\sum_{i<j}^{\availTestCaseNum-1}\testCase{i}\testCase{j}
\end{equation*}
\begin{example}
In the running example, the objective function for \num becomes $\objFunc_{\num}(\testCaseVector) = \testCase{0} + \testCase{1}+2\testCase{0}\testCase{1}$.
\terminateExample
\end{example}
The \textit{overall objective function} $\objFuncOverall$ can then be calculated based on the objectives corresponding to the properties and the objective of the number of selected test cases. We integrate all the objectives with the weighted-sum approach, where each objective 
is first normalized and then multiplied with a specific weight ($\weight{\num}$ or $\weight{\prop{h}}$) to reflect its priority in the TCM. For clarity, we do not show normalization in the rest of the calculation. The formulation is shown below:
%
\begin{equation*}
\objFuncOverall(\testCaseVector)=\weight{\num}\cdot\objFunc_{\num}(\testCaseVector)+\sum_{\prop{h}\in\propSet}\weight{\prop{h}}\cdot\objFunc_{\prop{h}}(\testCaseVector)
\end{equation*}
\begin{example}
In our running example, the overall objective is defined as follows:
\begin{equation*}
\begin{array}{l}
\objFuncOverall(\testCaseVector) = \weight{\num}\objFunc_{\num}(\testCaseVector) + \weight{\et}\objFunc_{\et}(\testCaseVector)+\weight{\fr}\objFunc_{\fr}(\testCaseVector)
\end{array}
\end{equation*}
\terminateExample
\end{example}

We then transfer the overall objective function into a QUBO model. Since all constant terms are combined to represent the offset of the energy system, and it doesn't affect the optimization process, we do not need to consider it in the QUBO formulation. The general objective function is shown below: (See Eq.~\ref{eq:qubo}):
\begin{align*}
\min \objFuncOverall(\testCaseVector) = & \sum_{i=0}^{\availTestCaseNum-1}
\left(
\weight{\num} + \sum_{\prop{h}\in\propSet}\weight{\prop{h}}
\left(
{{\propValue{i}{\prop{h}}}^2}-2\theoLimit{\prop{h}}
\right)
\right)
\cdot \testCase{i}\\
 & + \sum^{\availTestCaseNum-1}_{i<j}
 \left(
2\weight{\num}
+
2\sum_{\prop{h}\in\propSet}\weight{\prop{h}}\propValue{i}{\prop{h}}\propValue{j}{\prop{h}}
 \right)
 \cdot \testCase{i}\testCase{j}
\end{align*}
\noindent where $\weight{\num}+\sum_{\prop{h}\in\propSet}\weight{\prop{h}}({\propValue{i}{\prop{h}}}^2-2\theoLimit{\prop{h}})$ represents the linear coefficients (the diagonal terms in the weight matrix $Q$) and $2\weight{\num}+2\sum_{\prop{h}\in\propSet}\weight{\prop{h}}\propValue{i}{\prop{h}}\propValue{j}{\prop{h}}$ represents the quadratic coefficients (off-diagonal terms in $Q$).

\begin{example}
With the property set $\propSet=\{\fr, \et\}$, the expanded objective function without the constant term is
\begin{align*}
\objFuncOverall(\testCaseVector) & =
\left(
\weight{\num}+\sum_{\prop{h}\in\propSet}\weight{\prop{h}}({\propValue{0}{\prop{h}}}^2 - 2\theoLimit{\et}\propValue{0}{\et})
\right)
\testCase{0}\\
 & +
\left(
\weight{\num}+\sum_{\prop{h}\in\propSet}\weight{\prop{h}}({\propValue{1}{\prop{h}}}^2 - 2\theoLimit{\et}\propValue{1}{\et})
\right)
\testCase{1}\\
 & +
2\left(
\weight{\num}+\sum_{\prop{h}\in\propSet}\weight{\prop{h}}\propValue{0}{\prop{h}}\propValue{1}{\prop{h}}
\right)
\testCase{0}\testCase{1}
\end{align*}
We set all three weight values as $0.33$, and we obtain:
\begin{align*}
\objFuncOverall(\testCaseVector)=-0.0099\testCase{0}+0.396\testCase{1}+0.5412\testCase{0}\testCase{1}
\end{align*}
The objective function of the format with the $Q$ matrix is:
\begin{equation*}
\objFuncOverall(\testCaseVector)=\testCaseVector Q {\testCaseVector}^{\transpose}=\begin{bmatrix}
t_0\,t_1
\end{bmatrix} 
\begin{bmatrix}
-0.0099 & 0.5412\\
0 & 0.396 \\
\end{bmatrix} 
\begin{bmatrix}
t_0\\
t_1 
\end{bmatrix} 
\end{equation*}
\terminateExample
\end{example}

\section{Experiment Design}\label{sec:experiment_design}

We first present the research questions in Sect.~\ref{sec:RQs}, followed by the experimental setup in Sect.~\ref{sec:exp_setup}, and evaluation metrics and statistical tests in Sect.~\ref{sec:metricsAndTests}.

\subsection{Research Questions}\label{sec:RQs}
\begin{compactitem}
\item[\textbf{RQ1}]
Does the proposed QUBO formulation solve TCM problems effectively? 
 This RQ examines whether our QUBO formulation is effective for TCM by checking the results of SA. If yes, we can use the same QUBO formulation for QA. This check is necessary as we want to avoid any factor that may influence QA effectiveness that is not due to the formulation, such as hardware noise. Next, we compare \method with SA to determine if it performs similarly to SA in solving TCM to justify its use.  
\item[\textbf{RQ2}]
How is the effectiveness of \method compared with the QA process of not employing any sub-problem decomposition?
This RQ checks if QA with bootstrap sampling can outperform QA without decomposing in solving a small-scale TCM problem. Note that solving large problems with it is practically infeasible. 
\item[\textbf{RQ3}]
How does \method compare with an existing QA algorithm with a D-Wave built-in sub-problem decomposing?
This RQ checks whether \method's bootstrapping sampling strategy helps it to perform better than an existing QA algorithm from D-Wave with its own sub-problem decomposing process.

\item[\textbf{RQ4}]
How is the time efficiency of \method compared with the other three approaches?
This RQ aims to check the time cost of all approaches to obtain a comprehensive view of their overall time efficiency. 
\end{compactitem}

\subsection{Experimental Setup and Execution}\label{sec:exp_setup}

\noindent\textbf{Datasets.} We selected three real-world datasets to evaluate \method: \pa and \io from ABB Robotics Norway\footnote{http://new.abb.com/products/robotics} from~\cite{spieker2017reinforcement}, and \gs\footnote{https://code.google.com/archive/p/google-shared-dataset-of-test-suite-results/wikis/DataFields.wiki} from Google. Each test case in these datasets has property values related to ``execution time'' and ``failure rate''. Since we need test cases in the original test suite \originalTestSuite that have the possibility to trigger failures, we filtered out test cases whose failure rates are 0. Eventually, we employed 89, 287, and 1663 test cases from \pa, \gs, \io, respectively, and formed the three datasets for our experiment.

\noindent\textbf{Baselines}
To evaluate \method, we also implement our TCM formulation in QA in the following three optimization processes and hence form three baselines. 
%
%
\noindent \textit{(1) Simulated Annealing (SA).}
We use the dwave-neal framework from D-Wave as the SA baseline. It uses the QUBO formulation for optimization but with classical SA. 
\noindent \textit{(2) Vanilla QPU (VQ).}
We directly solve the TCM problems by running QA directly on D-Wave's QA hardware without decomposing.
\noindent \textit{(3) EIDQ.}
We implement \textit{EIDQ} as our baseline approach, which is tailored from the default hybrid QA algorithm from D-wave, Kerberos. This algorithm runs two classical branches (i.e., tabu search and (\textit{SA})) and a QPU branch in parallel and searches for an optimal solution by taking the best one from those produced by the three branches iteratively. The number of iterations is a hyper-parameter. In our experiment, we disabled the classical branches and we already have \textit{SA} for comparison. The QPU branch of Kerberos is interesting as it also combines a classical decomposing technique, i.e., Energy Impact Decomposer~\cite{energyimpact}, which forms a good comparison baseline. For convenience, we name the modified Kerberos as \textit{EIDQ}. The decomposing technique applies search to find a sub-problem of a certain size that maximally contribute to the overall objective~\cite{energyimpact}.

\noindent\textbf{Parameters.}
Each test case in each dataset has two properties: execution time and failure rate.
For failure rate, we calculated the upper limit ($\theoLimit{\fr}$) as the sum of the failure rates of all test cases in one dataset or one subset of it for \method since we aim to maximize the failure rate of the test suite. For execution time, we set the lowest limit (i.e., $\theoLimit{\et}$) to 0 since we aim to minimize the execution time of the test suite. We use equal weights in \objFuncOverall. Moreover, we used the default settings in the D-Wave framework for hardware execution. For the three QPU-based approaches (i.e., \method, \textit{EIDQ} and \textit{VQ}), we set the number of reads (see Sect.~\ref{sec:background}) as 100 in each execution. For \textit{SA}, we set the number of iterations as 100. The coverage rate $\beta$ in our experiment is 90\%.

We employed a range of sub-problem sizes \subproblemSize (defining the number of test cases in each sub-problem, see Sect.~\ref{sec:objective}) to conduct a series of experiments with \method. For \gs and \io, we went from 10 to 160 with an increment of 10 to identify the best \subproblemSize. We stopped at 160 since our pilot study results show that the current QA hardware was unstable in finding embeddings for more than 160 variables. For \pa, the total size of the dataset is 89; therefore, we use \subproblemSize from 10 to 80, with an increment of 10. For a fair comparison, we used the same number of variables (as for \method) as the maximum sub-problem size of \textit{EIDQ} in all iterations. We repeat experiments on each dataset 10 times for each sub-problem size in \method and \textit{EIDQ}. For \textit{SA} and \textit{VQ}, since there is no decomposing, we repeat experiments 10 times. 

\noindent\textbf{Execution Environment.} We used the D-Wave Advantage system~\cite{dwave} as the QPU, and the classical part of the experiment was run on one CPU node in the Ex3\footnote{EX3 is a national, experimental, heterogeneous computational cluster for researchers conducting experiments.} cluster with Intel(R) Xeon(R) Platinum 8168 CPU @ 2.70GHz.

\subsection{Evaluation Metrics and Statistical Tests}\label{sec:metricsAndTests}
\subsubsection{\textbf{Evaluation Metrics}}\label{subsec:metric}
We evaluate the approaches' effectiveness with the \fv metric (the lower, the better), which is related to the calculated objective function values. For \method and \textit{EIDQ}, for each sub-problem size, \fv for a TCM problem is calculated by taking the average value of the objective function (\objFuncOverall) values of the 10 optimal solutions found in 10 repetitions. For \textit{SA} and \textit{VQ}, as there are no sub-problems, \fv is calculated by taking the average value of the objective function (\objFuncOverall) values of the 10 optimal solutions found for 10 repetitions.
In addition, to compare \method with \textit{SA} (RQ1) and \textit{VQ} (RQ2), we evaluate their effectiveness with the \fv metric for each sub-problem size and also with the best sub-problem size. \method and \textit{EIDQ} are compared (RQ3) by various sub-problem sizes' effectiveness measured with \fv.
%
In RQ4, we evaluate the efficiency of \method in terms of the optimization time \ext. For \textit{SA} and \textit{VQ}, we calculate \ext as the sum of the classical CPU time for SA and QPU time for VQ. For \method and \textit{EIDQ}, \ext is the sum of the decomposing time and the QPU time. For \method (or \textit{EIDQ}), we calculate the bootstrap sampling time (or the decomposing time) and cumulative QPU portion time for all sub-problems. We used the best sub-problem sizes for both \method and \textit{EIDQ} to evaluate the total time cost and the number of qubits when comparing with \textit{SA} and \textit{VQ}. 


\subsubsection{\textbf{Statistical Tests}}\label{subsec:statisticaltests}
Following the guideline from~\cite{arcuri2011practical}, to establish the statistical significance of the results, we compare \method and \textit{EIDQ} with the \fv metrics (see Section~\ref{subsec:metric}) for all sub-problem sizes for each dataset using the Mann-Whitney U test and Vargha and Delaney's $\hat{A}_{12}$ effect size.
We set the significance level as 0.05. The null hypothesis is that there is no significant difference between the two approaches. If the null hypothesis is not rejected, we consider the effectiveness of the two approaches to be equal. Otherwise, if the null hypothesis is rejected, we utilize Vargha and Delaney's $\hat{A}_{12}$ statistic as the effect size measure to quantify the magnitude of the difference between the two groups. If $\hat{A}_{12}$ is 0.5, the result is achieved by chance. If $\hat{A}_{12}$ is smaller than 0.5, then \method obtains lower values than \textit{EIDQ} with a high probability and vice versa. For $\hat{A}_{12}$ statistics, the effect size in the range $(0.34, 0.44]$ and $[0.56, 0.64)$ are interpreted as \emph{Small}, $(0.29, 0.34]$ and $[0.64, 0.71)$ are interpreted as \emph{Medium}, and $[0, 0.29]$ and $[0.71, 1]$ are interpreted as \emph{Large}.

\section{Experiment Results}\label{sec:experimentResults}
In this section, we discuss experimental results to answer each RQ. The implementation of \method and all experiment results are provided in the online repository: \url{https://github.com/AsmarMuqeet/BootQA}.

\subsection{RQ1: Effectiveness of QUBO formulation}
We implement our QUBO formulation in \textit{SA} to evaluate its effectiveness in solving three TCM problems of the three datasets. We obtain \fv values of $1.19\text{e-}2$, $1.74\text{e-}4$, and $2.40\text{e-}2$ for \pa, \gs, and \io, respectively, which are all very close to 0 (smaller is better), implying that our QUBO formulation effectively guides SA to find optimal solutions for all three TCM problems. 

Encouraged by this, we then solve the TCM problems with \method with various sub-problem sizes (see Sect.~\ref{sec:exp_setup}). We calculated the \fv value for each sub-problem size and observed that for all three datasets, the \fv values obtained by \method with all sub-problem sizes are close to those produced by \textit{SA}; the average \fv value differences are $2.71\text{e-}3$, $2.63\text{e-}3$ and $1.10\text{e-}2$, for \pa, \gs and \io, respectively. When comparing \method configured with the best sub-problem sizes (30, 20, and 30 for \pa, \gs, and \io, respectively) with \textit{SA}, the average \fv differences are even more minor: $3.09\text{e-}4$, $8.21\text{e-}6$, and $7.04\text{e-}4$ for \pa, \gs, and \io, respectively. 
Thus, we can conclude that the effectiveness of \method is practically similar to \textit{SA}.

\subsection{RQ2: Effectiveness of \method compared to \textit{VQ}}

Large TCM problems cannot be directly solved on QA due to the limited number of qubits and couplers possible on the current QA hardware. For example, the overall QUBO graph scale of the two larger datasets (i.e., \gs and \io) exceeds the capacity of the current hardware. Thus, decomposing strategies are needed to fit a large TCM problem into the QA hardware.

To assess the effectiveness of using the bootstrap sampling strategy in QA to solve TCM problems, we compare the effectiveness of \method and \textit{VQ} for solving a small TCM problem with \pa. Recall from Sect.~\ref{sec:experiment_design} that we conduct experiments on \method configured with 8 different sub-problem sizes for \pa. 
We used Mann-Whitney U statistical test and Vargha and Delaney's $\hat{A}_{12}$ statistics to compare \objFuncOverall values produced by \method configured with each sub-problem size with those produced by \textit{VQ} in 10 repetitions. Results show that when \method is configured with 5 out of the 8 sub-problem sizes (i.e., 10, 20, 30 (the best sub-problem size), 40, and 50), the p-values are less than 0.05 with a large magnitude in favor of \method. No significant difference can be observed when \method has the 60 and 70 sub-problem sizes. For the sub-problem size 80, \textit{VQ} significantly outperforms \method with a large magnitude. 
Thus, we can conclude that \method's bootstrapping strategy with the best sub-problem size can potentially solve large TCM problems compared with implementing QA without decomposition.

\subsection{RQ3: Effectiveness of \method comparing to \textit{EIDQ}}\label{sec:rq3}
\begin{figure}[!tb]
\centering
\begin{subfigure}[b]{0.32\textwidth}
\centering
\includegraphics[width=\textwidth]{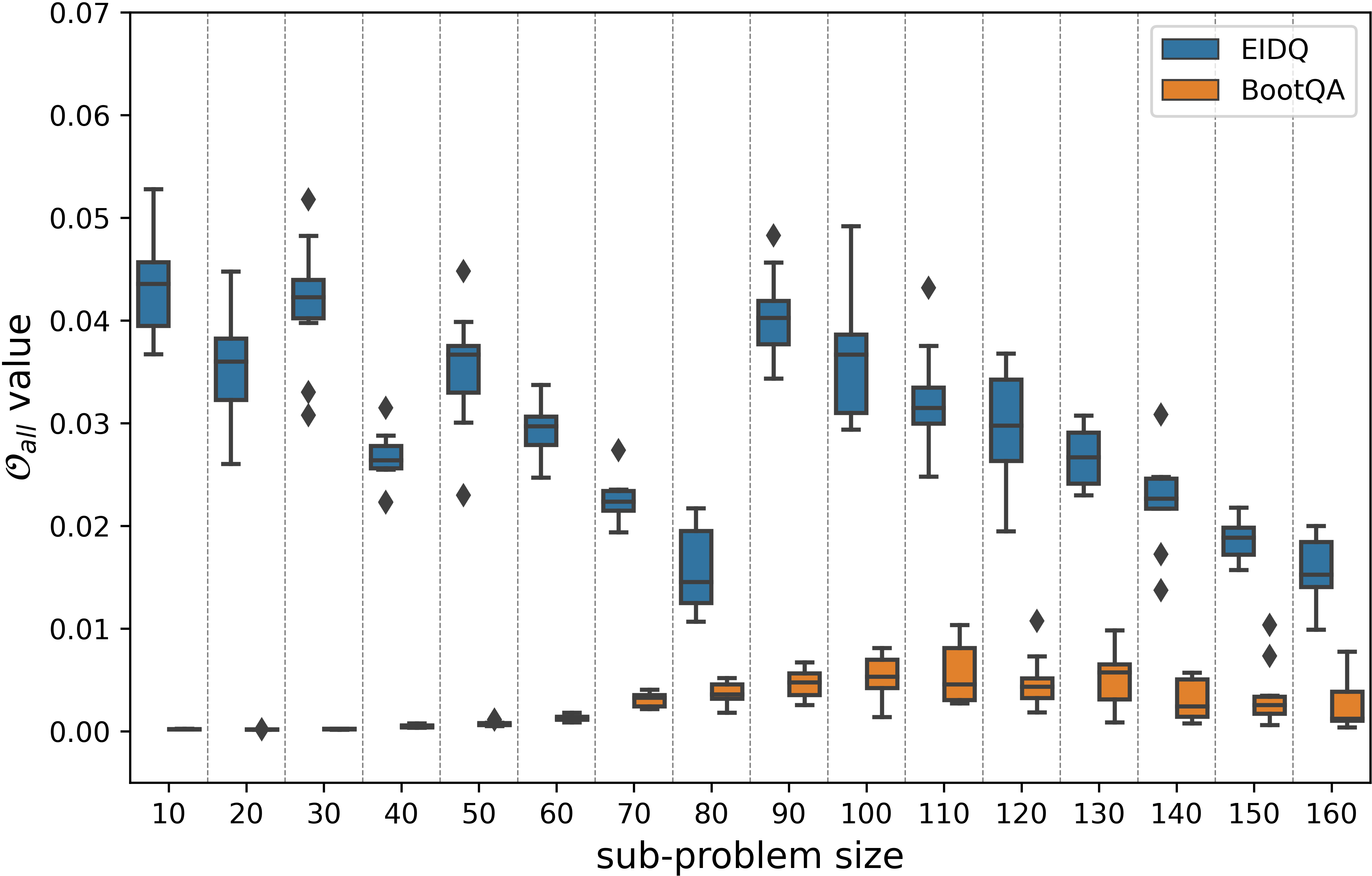}
\caption{\gs}
\label{fig:rq1_gsdtsr}
\end{subfigure}
\begin{subfigure}[b]{0.32\textwidth}
\centering
\includegraphics[width=\textwidth]{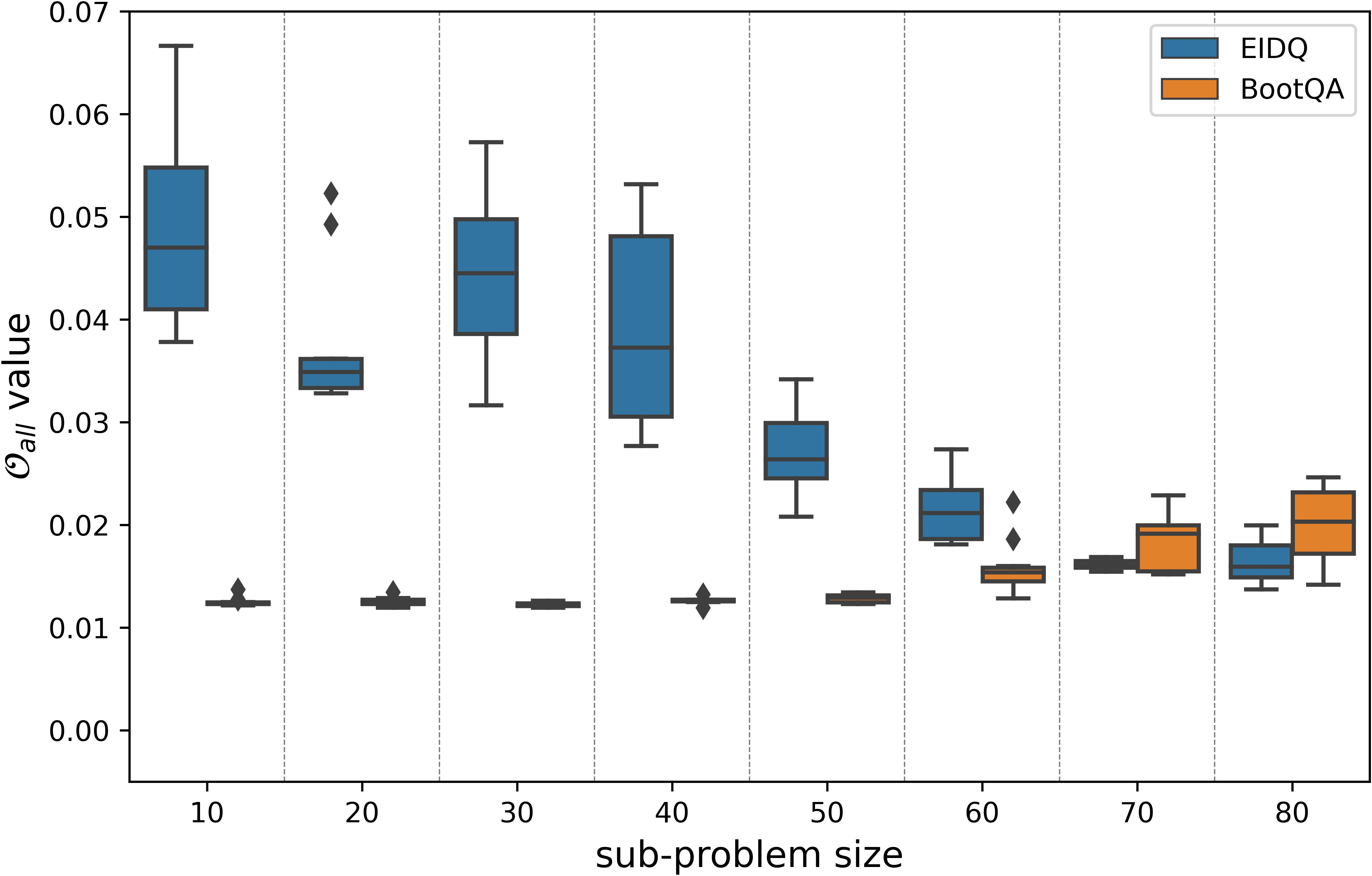}
\caption{\pa}
\label{fig:rq1_paintcontrol}
\end{subfigure}
\begin{subfigure}[b]{0.32\textwidth}
\centering
\includegraphics[width=\textwidth]{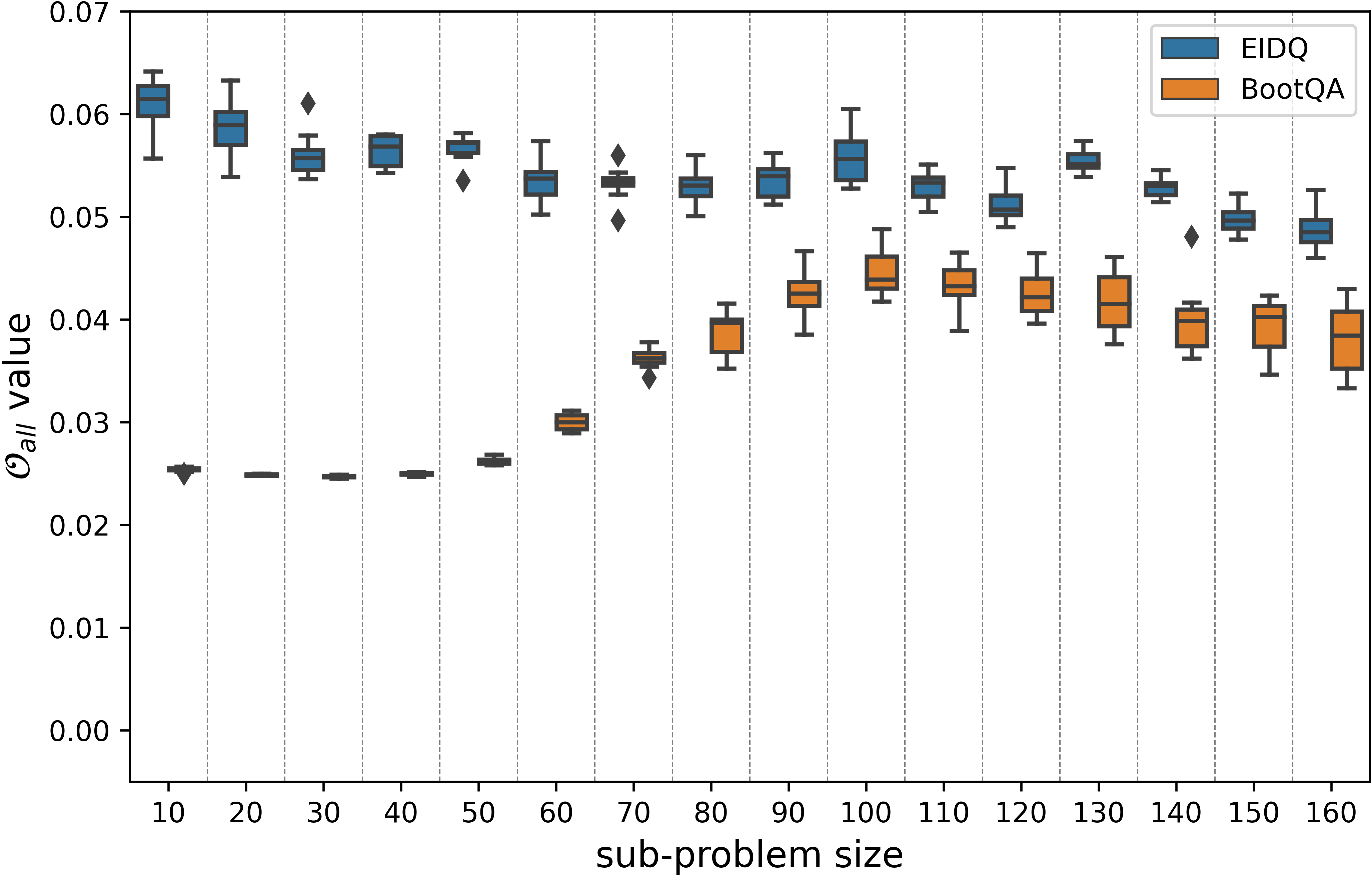}
\caption{\io}
\label{fig:rq1_iofrol}
\end{subfigure}
\caption{RQ3 -- Objective function (\objFuncOverall) values produced by \method and \textit{EIDQ} for each sub-problem}
\label{fig:fv_compare}
\end{figure}

Considering that both \method and \textit{EIDQ} solve a TCM problem by splitting it into sub-problems, their sizes (i.e., the number of test cases in each sub-problem) might impact their performance. According to Sect.~\ref{sec:exp_setup}, we employ a range of sub-problem sizes for both \method and \textit{EIDQ}. To compare them, for each sub-problem size, we analyze the objective function (\objFuncOverall) values of the optimal solutions produced by \method and \textit{EIDQ} in 10 repetitions. Results are shown in Fig.~\ref{fig:fv_compare}.
Comparing the three datasets, we can see that, in most cases, the objective function values of \method are lower than that of \textit{EIDQ}, except for the cases with the 70 and 80 sub-problem sizes in \pa. Moreover, one can observe that for all three datasets, \method produced solutions with less variability among different sample sizes compared to \textit{EIDQ}. \method follows a trend of having lower objective function values when the sub-problem size is small and higher objective function values as the sub-problem size grows. In the case of \textit{EIDQ}, no definite trend with respect to the sub-problem size can be observed. For instance, in \gs, sometimes the objective function values are comparatively high even with small sample sizes like 10 and 30. 

We find that \method consistently achieves lower objective function values with smaller sub-problem sizes (e.g., from 10 to 50), while larger objective function values for larger sub-problem sizes, as shown in Fig.~\ref{fig:fv_compare}. One possible reason is that for a larger sub-problem, the length of the qubit chain will be longer, which affects the QA accuracy~\cite{abbott2019hybrid, Calaza2021}. In addition, for D-Wave QA, all coefficients in the QUBO formulation need to be scaled down to fit the fixed hardware bias and couplers determined by the physical properties of the qubits and their interconnections~\cite{Calaza2021}. 
For a larger sub-problem (with more variables), the effective energy gap\footnote{Effective energy gap is the difference of energy between the ground state and the first excited state of the quantum system~\cite{energygap}.} becomes smaller. This is because the problem Hamiltonian gets more complex and has more energy levels, which results in a denser spectrum of energy levels and a smaller effective energy gap, consequently leading to lower optimization performance~\cite{Calaza2021,techqpu}.

By observing the trend of objective function values generated by \method in Fig.~\ref{fig:fv_compare}, we can observe a rise of the objective function values in \gs and \io at the sub-problem size being 110 and 100, respectively. The possible reason is that the scaling effect on different QUBO formulations varies, consequently affecting the results for different sub-problems.

Overall, in most cases for all datasets, \method performs better than \textit{EIDQ}. To establish the statistical significance of the results, we also evaluated the values of \fv for all sub-problem sizes of \method and \textit{EIDQ} with the Mann-Whitney U statistical test and Vargha and Delaney's $\hat{A}_{12}$ statistics. For all three datasets, the p-values are all less than 0.05, showing a significant difference between \method and \textit{EIDQ}. For \gs and \io, the effect sizes are 0 (less than 0.5), indicating that \method is significantly better than \textit{EIDQ} with a large magnitude for all sub-problem sizes. For the \pa dataset, the effect size is 0.062, which is also less than 0.5, showing that \method significantly performs better than \textit{EIDQ} with a large magnitude in most sub-problem sizes. 

We also compare \method and \textit{EIDQ} with their best sub-problem sizes for each dataset, regarding the \fv values. Results are summarized in Table.~\ref{table:best}. \method gets the best results with the sub-problem sizes of $\subproblemSize=20$ for \gs and $\subproblemSize=30$ for \io and \pa. For \textit{EIDQ}, the best settings are 70 for \pa, 160 for \gs and \io. The lowest \fv produced by \method in \pa, \gs, and \io are $1.22\text{e-}2$, $1.82\text{e-}4$ and $2.46\text{e-}2$, respectively, while that for \textit{EIDQ} are $1.62\text{e-}2$, $1.55\text{e-}2$ and $4.87\text{e-}2$. Comparing the two approaches, \method gets better performance in all datasets.
\begin{table}[!tb]
\small
\centering
\caption{Best sub-problem sizes of \method and \textit{EIDQ}. \subproblemSize is the sub-problem size; \numOfSampledSubsets denotes the number of subsets \method used for reaching 90\% coverage; $itr$: the average number of iterations \textit{EIDQ} used to produce the final solution.}
\label{table:best}
\renewcommand\arraystretch{1.5}
\setlength{\tabcolsep}{2.5pt}
\resizebox{0.35\textwidth}{!}{
\begin{tabular}{cc|cc|cc}
\hline
\multicolumn{2}{c}{Dataset} & \multicolumn{2}{c}{\method} & \multicolumn{2}{c}{$EIDQ$}\\
\hline
Name&Size & \subproblemSize & \numOfSampledSubsets & \subproblemSize & $itr$ \\
\hline
\pa &89& 30 & 6 & 70 & 16 \\
\gs &287& 20 & 21 & 160 & 7 \\
\io &1663& 30 & 127 & 160 & 4 \\
\hline
\end{tabular}
}
\end{table}

\subsection{RQ4: Efficiency of \method}
To compare the time efficiency, \method and \textit{EIDQ} are set with their best sub-problem sizes. For \textit{SA}, we calculate the computation time of the whole optimization process as \ext. For \method, \textit{EIDQ}, and \textit{VQ}, we employ the ``QPU access time'' metric~\cite{techqpu, abbott2019hybrid} to measure the QPU portion in \ext, which is obtained via the D-Wave QPU-API. Specifically, the QPU access time measures the total time spent on Step 3 (programming, initialization, and sampling) presented in Sect.~\ref{sec:background}. 
For \method and \textit{EIDQ}, we use the time required for bootstrap sampling and Energy Impact Decomposer (see Sect.~\ref{sec:exp_setup}) as their decomposing time, respectively. 
Results are shown in Table~\ref{table:time_four}.

\begin{table*}[]
\large
\centering
\caption{RQ4 -- Time performance (in seconds) of the four approaches. Each value is shown in the format of \textit{average} ± \textit{standard deviation}. \ext values highlighted in the blue background are the smallest among all approaches in one dataset.}
\label{table:time_four}
\renewcommand\arraystretch{1.4}
\resizebox{\columnwidth}{!}{
\begin{tabular}{cccccccccc}
\hline
Dataset                           & \multicolumn{1}{c|}{\textit{SA}}                                            & \multicolumn{3}{c|}{\textit{BootQA}}                                                                                                              & \multicolumn{3}{c|}{\textit{EIDQ}}                                                                                                                & \multicolumn{2}{c}{\textit{VQ}}                                                               \\ \hline
\multicolumn{1}{c|}{}                      & \multicolumn{1}{c|}{}                                  & \multicolumn{2}{c|}{\textit{exTime}}                                                 & \multicolumn{1}{c|}{}                                     & \multicolumn{2}{c|}{\textit{exTime}}                                                 & \multicolumn{1}{c|}{}                                     & \multicolumn{1}{c|}{}                                  &                                      \\ \cline{3-4} \cline{6-7}
\multicolumn{1}{c|}{\multirow{-2}{*}{}}    & \multicolumn{1}{c|}{\multirow{-2}{*}{\textit{exTime}}} & \multicolumn{1}{c|}{\textit{Decomposing}} & \multicolumn{1}{c|}{\textit{QPU access}} & \multicolumn{1}{c|}{\multirow{-2}{*}{\textit{Embedding}}} & \multicolumn{1}{c|}{\textit{Decomposing}} & \multicolumn{1}{c|}{\textit{QPU access}} & \multicolumn{1}{c|}{\multirow{-2}{*}{\textit{Embedding}}} & \multicolumn{1}{c|}{\multirow{-2}{*}{\textit{exTime}}} & \multirow{-2}{*}{\textit{Embedding}} \\ \hline
\multicolumn{1}{c|}{\textit{PaintControl}} & 0.440s ± 0.005                                         & \textless{}0.001                          & 0.171s ± 0.006                           & 40.725s ± 6.019                                           & 1.542s ± 2.333                            & 0.531s ± 0.411                           & 641.678s ± 529.882                                        & \cellcolor[HTML]{ECF4FF}0.036s ± 0.001                 & 76.091s ± 15.516                     \\
\multicolumn{1}{c|}{\textit{GSDTR}}        & 2.811s ± 0.021                                         & \cellcolor[HTML]{ECF4FF}\textless{}0.001  & \cellcolor[HTML]{ECF4FF}0.890s ± 0.009   & 179.665s ± 10.211                                         & 1.438s ± 0.031                            & 0.167s ± 0.001                           & 1998.118s ± 385.059                                       & -                                                      & -                                    \\
\multicolumn{1}{c|}{\textit{IOFRL}}        & 86.333s ± 0.602                                        & \cellcolor[HTML]{ECF4FF}\textless{}0.001  & \cellcolor[HTML]{ECF4FF}3.656s ± 0.024   & 994.821s ± 10.245                                         & 70.726s ± 0.427                           & 0.293s ± 0.002                           & 3459.223s ± 730.282                                       & -                                                      & -                                    \\ \hline
\end{tabular}
}
\end{table*}

Regarding \ext, comparing all four approaches, for the \pa dataset, since \textit{VQ} takes only one execution to solve the problem, its \ext is the shortest while \textit{EIDQ}'s \ext (the sum of the decomposing time and the QPU access time) is the longest. For \gs and \io, \method's \ext is the shortest among all approaches except for \textit{VQ} as it only applies to \pa. 

When looking at each approach's time performance across the three datasets of various sizes, we can observe from Table \ref{table:time_four} that \ext of \textit{SA} grows significantly as the dataset size increases. Notably, the dataset size growth from \gs to \io is around 5.8 times ($1663/287$), whereas \textit{SA}'s \ext surges around 30 times ($86.333/2.811$). For \textit{EIDQ}, we observe that its \textit{Decomposing} time takes a large part of \ext, which mainly depends on the number of iterations and the sub-problem size. The QPU access time of \textit{EIDQ} is very short, on the other hand. Moreover, the standard deviation of its decomposing time varies more than the \method, which reflects the influence of the number of iterations on the decomposing time. 
For \method, the decomposing time of \method (i.e., bootstrap sampling time) is negligible for all datasets (i.e., less than 0.001 seconds), and the main time cost of \method is the QPU access time, which is affected by the problem size and the times of sampling \numOfSampledSubsets. 

When comparing the time performance of all four approaches across all three datasets, \ext of \textit{SA} and the decomposing time of \textit{EIDQ} (i.e., the classical portion of \ext) grows dramatically with the increase of dataset size, which is, however, not the case for \method, where the decomposing time increases from 0.171s only to 3.656s. Thus, for large datasets, \textit{SA} and some hybrid QA (e.g., \textit{EIDQ}) may require considerable time to produce solutions. Comparatively, \method serves as a more time-efficient method for large TCM problems.

Regarding time for embedding, as discussed in Sect.~\ref{sec:background}, the general D-Wave QA process contains a minor-embedding phase, during which the heuristic-based D-Wave Advantage system~\cite{dwave} calculates an approximate mapping from a QUBO graph onto a quantum hardware working graph. This means the embedding can differ in each run, even for the same QUBO graph on the same hardware. 
As shown in Table~\ref{table:time_four}, the embedding time used by \method is the shortest as compared to \textit{EIDQ} and \textit{VQ}. Generally, the embedding time grows with the increase in the size of the QUBO graph~\cite{choi2008minor}. However, we would like to mention that to compare with \textit{SA}, though \method requires a significant amount of time for embedding, it is feasible to reuse the embedding across sub-problems, as we will discuss in detail in Sect. \ref{discussions}. Therefore, the overall cost of applying \method can be further reduced. Optimizing the time and effectiveness of embedding is an active area of research~\cite{emb1,emb2,emb3,emb4}. In the near future, the possibility of having a general embedding for fully connected QUBO graphs that can be used to run multiple problems in parallel will reduce the embedding cost drastically~\cite{parallel}.

\subsection{Threats to Validity}\label{sec:threats}
There are several threats to validity~\cite{Wohlin2012}. We evaluated our approach on only three datasets due to the limitation of hardware resources. However, the three chosen datasets are industrial case studies from Google and ABB Robotics. Their sizes range from 89 to 1663, which helps to evaluate the approaches with problems of various scales. In addition, conducting experiments with more datasets is constrained by the practical limitation on the QA hardware access, which affects the scale of the experiment. 

Regarding the QA hyper-parameters configuration, we used the default settings from D-Wave in our experiment. Though we are aware that D-Wave QA solver instances have several hyper-parameters\footnote{D-Wave solver parameter list:  \url{https://docs.dwavesys.com/docs/latest/c_solver_parameters.html}}, different settings of which may have an impact on the QA performance, empirically identifying the best settings of the hyper-parameters is very costly, which we cannot afford for now. 

Another threat to the empirical evaluation is the metrics used for answering RQs. We used the same fitness function for all approaches to evaluate their effectiveness. We use the experiment's average values of 10 repetitions to ensure accurate results. For calculating the time efficiency of the approaches, we used the standard metric, the QPU access time~\cite{techqpu, abbott2019hybrid}. To establish the statistical significance of the results, we employed the Mann-Whitney U and Vargha-Daleney's $\hat{A}_{12}$ statistics to verify our findings. With the above, we provide a fair evaluation of all approaches and reduce the randomness during experiments.

\section{Discussion}\label{discussions}

\subsection{Enriching the QUBO formulation with constraints.} Our results showed that for all three datasets, \method performs better in terms of \fv for smaller sub-problem size (i.e., \subproblemSize) since larger problem sizes require more chains and couplers in the working graph. Moreover, scaling is required to fit the problem into the physical hardware. Though \method performs better with smaller \subproblemSize, it increases the number of \numOfSampledSubsets as all sub-problems together need to cover a required percentage of the dataset (e.g., 90\%). A large \numOfSampledSubsets increases the overall execution time of QA, which might limit applying \method in cases where a solution is needed within a timing deadline. One solution is to optimize the QUBO formulation by considering dependencies (e.g., as constraints) among test cases (e.g., one test case needs to be executed before another). Our current QUBO formulation for the TCM problem makes a fully connected QUBO graph that cannot directly map to the QA hardware without using chains (see Section~\ref{sec:background}). Considering dependencies among test cases during the QUBO formulation could help make the QUBO graph sparsely connected, which can improve QA performance by requiring smaller chains and coupler strengths \cite{abbott2019hybrid}. Also, with sparsely connected graphs, more variables can be mapped directly to physical hardware, which allows the use of large sub-problem size \subproblemSize without reducing the effectiveness of the approach.
\subsubsection{Reusing the embedding across sub-problems.}
The minor-embedding phase is the most time-consuming part of the whole QA process, as we observed from the RQ4 results. A good embedding can significantly improve the accuracy and time performance of QA~\cite{abbott2019hybrid}. To solve a TCM problem with \method, the sub-problem size \subproblemSize is fixed, and given a fully connected QUBO graph, one can easily reuse the embedding of one sub-problem to solve a new sub-problem. One possible solution to find a good embedding is to use search algorithms to find a good embedding of a fixed-size fully connected QUBO graph. Such an embedding can be used for both fully and sparsely connected QUBO graphs as long as the number of nodes in the graph remains the same. Reusing embedding can significantly improve the time performance of QA, hence making it applicable in real settings even with the currently limited size of quantum hardware.

\subsubsection{Hardware resource usage.}
Current QA hardware has a limited number of qubits, which limits the size of optimization problems that can be solved with QA. However, from the results of RQ1, we observed that \method performs comparably to SA by dividing a bigger problem into smaller sub-problems. Table
~\ref{tab:qubit_count} shows the average number of physical qubits ($\nq$) used for solving the sub-problems with various sizes throughout 10 repetitions with \method and \textit{EIDQ}. 
\begin{table}[!tb]
\centering
\caption{Average number of physical qubits (\nq) used by \method and \textit{EIDQ} for solving all TCM problems and obtained \fv values. \nq required and \fv values produced by \method and EIDQ with the best sub-problem sizes are highlighted in bold.}
\label{tab:qubit_count}
\renewcommand\arraystretch{1.1}
\resizebox{\columnwidth}{!}{
\begin{tabular}{c|cccccccccccc}
\hline
\subproblemSize   & \multicolumn{12}{c}{Dataset}                                                                                                                                                                                                                                                                                  \\ \hline
\multirow{3}{*}{} & \multicolumn{4}{c|}{\pa}                                                                              & \multicolumn{4}{c|}{\gs}                                                                              & \multicolumn{4}{c}{\io}                                                                       \\ \cline{2-13} 
                  & \multicolumn{2}{c|}{\method}                       & \multicolumn{2}{c|}{\textit{EIDQ}}                & \multicolumn{2}{c|}{\method}                       & \multicolumn{2}{c|}{\textit{EIDQ}}                & \multicolumn{2}{c|}{\method}                       & \multicolumn{2}{c}{\textit{EIDQ}}         \\ \cline{2-13} 
                  & \multicolumn{1}{c|}{\nq} & \multicolumn{1}{c|}{\fv} & \multicolumn{1}{c|}{\nq} & \multicolumn{1}{c|}{\fv} & \multicolumn{1}{c|}{\nq} & \multicolumn{1}{c|}{\fv} & \multicolumn{1}{c|}{\nq} & \multicolumn{1}{c|}{\fv} & \multicolumn{1}{c|}{\nq} & \multicolumn{1}{c|}{\fv} & \multicolumn{1}{c|}{\nq} & \fv              \\ \hline
10                & 17.3                    & 0.0125                  & 17.2                    & 0.0490                  & 17.2                    & 0.0002                  & 17.2                    & 0.0432                  & 17.2                    & 0.0253                  & 17.2                    & 0.0610          \\
20                & \cellcolor[HTML]{ECF4FF}\textbf{55.8}           & \cellcolor[HTML]{ECF4FF}\textbf{0.0125}         & 55.9                    & 0.0376                  & \cellcolor[HTML]{ECF4FF}\textbf{55.3}           & \cellcolor[HTML]{ECF4FF}\textbf{0.0001}         & 55.8                    & 0.0355                  & \cellcolor[HTML]{ECF4FF}\textbf{55.5}           & \cellcolor[HTML]{ECF4FF}\textbf{0.0248}         & 55.6                    & 0.0587          \\
30                & \cellcolor[HTML]{ECF4FF}\textbf{123.5}          & \cellcolor[HTML]{ECF4FF}\textbf{0.0122}         & 122.9                   & 0.0443                  & \cellcolor[HTML]{ECF4FF}\textbf{122.4}          & \cellcolor[HTML]{ECF4FF}\textbf{0.0002}         & 123.4                   & 0.0417                  & \cellcolor[HTML]{ECF4FF}\textbf{122.4}          & \cellcolor[HTML]{ECF4FF}\textbf{0.0247}         & 122.5                   & 0.0560          \\
40                & 212.1                   & 0.0126                  & 212.5                   & 0.0389                  & 213.7                   & 0.0005                  & 213.4                   & 0.0264                  & 214.4                   & 0.0249                  & 214.5                   & 0.0564          \\
50                & 326.5                   & 0.0128                  & 331.9                   & 0.0272                  & 328.3                   & 0.0007                  & 328.4                   & 0.0353                  & 330.3                   & 0.0262                  & 330.6                   & 0.0566          \\
60                & 468.2                   & 0.0158                  & 470.9                   & 0.0216                  & 465.0                   & 0.0012                  & 471.6                   & 0.0294                  & 467.1                   & 0.0300                  & 468.4                   & 0.0534          \\
70                & 621.1                   & 0.0183                  & \cellcolor[HTML]{ECF4FF}\textbf{629.8}          & \cellcolor[HTML]{ECF4FF}\textbf{0.0161}         & 623.8                   & 0.0030                  & 633.5                   & 0.0226                  & 627.5                   & 0.0362                  & 624.3                   & 0.0532          \\
80                & 817.2                   & 0.0201                  & 821.7                   & 0.0164                  & 827.3                   & 0.0037                  & 813.1                   & 0.0156                  & 823.3                   & 0.0387                  & 826.5                   & 0.0530          \\
90                & -                       & -                       & -                       & -                       & 1045.2                  & 0.0046                  & 1032.1                  & 0.0404                  & 1040.7                  & 0.0425                  & 1055.6                  & 0.0536          \\
100               & -                       & -                       & -                       & -                       & 1274.0                  & 0.0053                  & 1270.0                  & 0.0365                  & 1279.2                  & 0.0446                  & 1284.1                  & 0.0558          \\
110               & -                       & -                       & -                       & -                       & 1560.4                  & 0.0055                  & 1535.4                  & 0.0323                  & 1567.0                  & 0.0432                  & 1560.3                  & 0.0530          \\
120               & -                       & -                       & -                       & -                       & 1882.2                  & 0.0048                  & 1884.0                  & 0.0298                  & 1887.1                  & 0.0424                  & 1854.0                  & 0.0512          \\
130               & -                       & -                       & -                       & -                       & 2160.1                  & 0.0051                  & 2282.7                  & 0.0265                  & 2223.8                  & 0.0416                  & 2286.5                  & 0.0553          \\
140               & -                       & -                       & -                       & -                       & 2610.9                  & 0.0030                  & 2589.5                  & 0.0224                  & 2632.5                  & 0.0399                  & 2619.2                  & 0.0528          \\
150               & -                       & -                       & -                       & -                       & 3038.6                  & 0.0034                  & 3143.7                  & 0.0186                  & 3020.8                  & 0.0393                  & 3043.9                  & 0.0497          \\
160               & -                       & -                       & -                       & -                       & 3475.0                  & 0.0025                  & \cellcolor[HTML]{ECF4FF}\textbf{3406.8}         & \cellcolor[HTML]{ECF4FF}\textbf{0.0155}         & 3462.1                  & 0.0381                  & \cellcolor[HTML]{ECF4FF}\textbf{3406.7}         & \cellcolor[HTML]{ECF4FF}\textbf{0.0487} \\ \hline
\end{tabular}
}
\end{table}

As shown in Table.~\ref{tab:qubit_count}, for each \subproblemSize, $\nq$ used by both approaches is comparable. And, it grows dramatically with the increase of \subproblemSize because the length of qubit chains grows with the increase in the number of variables. 
Specifically, for \textit{VQ} with all 89 variables in \pa, the number of qubits used is around 1000, which is enormous compared to the size of the problem. In contrast, \method can get relatively good performance with a small number of qubits; around 56 qubits were used when \subproblemSize was set 20, and 123 qubits were used for \subproblemSize set 30. For \textit{EIDQ}, to get a good performance, as compared to \method, a lot more qubits are required. For instance, around 630 (3407) qubits were needed for \subproblemSize being 70 (160). Thus, for solving TCM problems with a good performance by QA, the requirement of hardware for \method is not as high as that for other approaches, making \method a potentially more resource-efficient approach for solving these types of problems. In addition, we observe that with the current hardware, using more qubits does not guarantee good performance and the reasons are explained in RQ3 (Sect.~\ref{sec:rq3})

\section{Conclusion and Future Work}\label{sec:conclusions}
Quantum annealers aim to solve combinatorial optimization problems, including those in software engineering, such as test optimization. In this paper, we present the first work in the literature that provides a generic formulation of the test case minimization (TCM) problem for quantum annealing (QA). Moreover, given the fact that the current quantum annealers have a limited number of quantum bits (qubits), we propose a sampling method to optimize the use of physical qubits for TCM problems. We implemented our TCM formulation in four optimization processes: classical simulated annealing (SA), QA without sampling (named \textit{VQ}), and QA with two different sampling methods (\method we newly proposed in this paper and \textit{EIDQ} tailored from D-Wave's Kerberos). We evaluate them on three real-world datasets of various scales. Results show that \method performs similarly as \textit{SA}, and outperforms both \textit{VQ} and \textit{EIDQ} in terms of effectiveness. \method also shows higher time efficiency when solving large-scale TCM problems compared with the other three. 
Our future work includes optimizing the QUBO formulation to further improve the efficiency of \method. We also plan to conduct more experiments with larger datasets.

\bibliographystyle{IEEEtran}

\begin{thebibliography}{10}
\providecommand{\url}[1]{#1}
\csname url@samestyle\endcsname
\providecommand{\newblock}{\relax}
\providecommand{\bibinfo}[2]{#2}
\providecommand{\BIBentrySTDinterwordspacing}{\spaceskip=0pt\relax}
\providecommand{\BIBentryALTinterwordstretchfactor}{4}
\providecommand{\BIBentryALTinterwordspacing}{\spaceskip=\fontdimen2\font plus
\BIBentryALTinterwordstretchfactor\fontdimen3\font minus
  \fontdimen4\font\relax}
\providecommand{\BIBforeignlanguage}[2]{{%
\expandafter\ifx\csname l@#1\endcsname\relax
\typeout{** WARNING: IEEEtran.bst: No hyphenation pattern has been}%
\typeout{** loaded for the language `#1'. Using the pattern for}%
\typeout{** the default language instead.}%
\else
\language=\csname l@#1\endcsname
\fi
#2}}
\providecommand{\BIBdecl}{\relax}
\BIBdecl

\bibitem{speed}
\BIBentryALTinterwordspacing
AY.~ Kitaev,  A.~Shen, MN.~Vyalyi,
``Classical and quantum computing,``
\emph{NY: Springer New York}, pp. 203--217, 2007. [Online]. Available:
  \url{https://doi.org/10.1007/978-0-387-36944-0_13}
\BIBentrySTDinterwordspacing

\bibitem{siloi2021investigating}
I.~Siloi, V.~Carnevali, B.~Pokharel, M.~Fornari, and R.~Di~Felice,
  ``Investigating the chinese postman problem on a quantum annealer,''
  \emph{Quantum Machine Intelligence}, vol.~3, no.~1, p.~3, 2021.

\bibitem{perdomo2015quantum}
A.~Perdomo-Ortiz, J.~Fluegemann, S.~Narasimhan, R.~Biswas, and V.~N.
  Smelyanskiy, ``A quantum annealing approach for fault detection and diagnosis
  of graph-based systems,'' \emph{The European Physical Journal Special
  Topics}, vol. 224, pp. 131--148, 2015.

\bibitem{inoue2020model}
D.~Inoue and H.~Yoshida, ``Model predictive control for finite input systems
  using the {D-Wave} quantum annealer,'' \emph{Scientific Reports}, vol.~10,
  no.~1, p. 1591, 2020.

\bibitem{farhi2000quantum}
E.~Farhi, J.~Goldstone, S.~Gutmann, and M.~Sipser, ``Quantum computation by
  adiabatic evolution,'' \emph{arXiv preprint quant-ph/0001106}, 2000.

\bibitem{Survey1}
\BIBentryALTinterwordspacing
R.~Mukherjee and K.~S. Patnaik, ``A survey on different approaches for software
  test case prioritization,'' \emph{J. King Saud Univ. Comput. Inf. Sci.},
  vol.~33, no.~9, pp. 1041--1054, nov 2021. [Online]. Available:
  \url{https://doi.org/10.1016/j.jksuci.2018.09.005}
\BIBentrySTDinterwordspacing

\bibitem{survey2}
\BIBentryALTinterwordspacing
R.~Pan, M.~Bagherzadeh, T.~A. Ghaleb, and L.~Briand, ``Test case selection and
  prioritization using machine learning: A systematic literature review,''
  \emph{Empirical Softw. Engg.}, vol.~27, no.~2, mar 2022. [Online]. Available:
  \url{https://doi.org/10.1007/s10664-021-10066-6}
\BIBentrySTDinterwordspacing

\bibitem{nsga}
S.~Verma, M.~Pant, and V.~Snasel, ``A comprehensive review on nsga-ii for
  multi-objective combinatorial optimization problems,'' \emph{Ieee Access},
  vol.~9, pp. 57\,757--57\,791, 2021.

\bibitem{ARRIETA2019137}
\BIBentryALTinterwordspacing
A.~Arrieta, S.~Wang, U.~Markiegi, A.~Arruabarrena, L.~Etxeberria, and
  G.~Sagardui, ``Pareto efficient multi-objective black-box test case selection
  for simulation-based testing,'' \emph{Information and Software Technology},
  vol. 114, pp. 137--154, 2019. [Online]. Available:
  \url{https://www.sciencedirect.com/science/article/pii/S0950584918301721}
\BIBentrySTDinterwordspacing

\bibitem{panichella2014improving}
A.~Panichella, R.~Oliveto, M.~Di~Penta, and A.~De~Lucia, ``Improving
  multi-objective test case selection by injecting diversity in genetic
  algorithms,'' \emph{IEEE Transactions on Software Engineering}, vol.~41,
  no.~4, pp. 358--383, 2014.

\bibitem{wang2015cost}
\BIBentryALTinterwordspacing
S.~Wang, S.~Ali, and A.~Gotlieb, ``Cost-effective test suite minimization in
  product lines using search techniques,'' \emph{J. Syst. Softw.}, vol. 103,
  no.~C, pp. 370--391, may 2015. [Online]. Available:
  \url{https://doi.org/10.1016/j.jss.2014.08.024}
\BIBentrySTDinterwordspacing

\bibitem{bootstrap}
A.~Kulesa, M.~Krzywinski, P.~Blainey, and N.~Altman, ``Sampling distributions
  and the bootstrap,'' 2015.

\bibitem{adiabatic_theorem}
M.~Born and V.~Fock, ``{Beweis des Adiabatensatzes},'' \emph{Zeitschrift
  f{\"{u}}r Physik}, no.~6, 1928.

\bibitem{QA_theory}
S.~Morita and H.~Nishimori, ``{Mathematical foundation of quantum annealing},''
  \emph{Journal of Mathematical Physics}, vol.~49, no.~12, 2008.

\bibitem{Yarkoni_2022}
\BIBentryALTinterwordspacing
S.~Yarkoni, E.~Raponi, T.~Bäck, and S.~Schmitt, ``Quantum annealing for
  industry applications: introduction and review,'' \emph{Reports on Progress
  in Physics}, vol.~85, no.~10, p. 104001, sep 2022. [Online]. Available:
  \url{https://dx.doi.org/10.1088/1361-6633/ac8c54}
\BIBentrySTDinterwordspacing

\bibitem{kadowaki1998quantum}
T.~Kadowaki and H.~Nishimori, ``Quantum annealing in the transverse {Ising}
  model,'' \emph{Physical Review E}, vol.~58, no.~5, p. 5355, 1998.

\bibitem{QUBO}
M.~Lewis and F.~Glover, ``Quadratic unconstrained binary optimization problem
  preprocessing: Theory and empirical analysis,'' \emph{Networks}, vol.~70,
  no.~2, pp. 79--97, 2017.

\bibitem{techqpu}
\BIBentryALTinterwordspacing
{D-Wave Systems Inc.}, ``Technical description of the d-wave quantum processing
  unit,'' May 2021. [Online]. Available:
  \url{https://docs.dwavesys.com/docs/latest/doc\_qpu.html}
\BIBentrySTDinterwordspacing

\bibitem{techqpu2020}
\BIBentryALTinterwordspacing
C.~McGeoch and P.~Farre, ``The {D-Wave} advantage systems: An overview,''
  {D-Wave} Systems Inc., Tech. Rep., 2020. [Online]. Available:
  \url{https://www.dwavesys.com/media/s3qbjp3s/14-1049a-a_the_d-wave_advantage_system_an_overview.pdf}
\BIBentrySTDinterwordspacing

\bibitem{Choi2008}
\BIBentryALTinterwordspacing
V.~Choi, ``Minor-embedding in adiabatic quantum computation: I. the parameter
  setting problem,'' \emph{Quantum Information Processing}, vol.~7, pp.
  193--209, 2008. [Online]. Available:
  \url{https://doi.org/10.1007/s11128-008-0082-9}
\BIBentrySTDinterwordspacing

\bibitem{rosenberg2016solving}
G.~Rosenberg, P.~Haghnegahdar, P.~Goddard, P.~Carr, K.~Wu, and M.~L. de~Prado,
  ``Solving the optimal trading trajectory problem using a quantum annealer,''
  \emph{IEEE Journal of Selected Topics in Signal Processing}, vol.~10, no.~6,
  pp. 1053--1060, 2016.

\bibitem{kulkarni2021quantum}
V.~Kulkarni, M.~Kulkarni, and A.~Pant, ``Quantum computing methods for
  supervised learning,'' \emph{Quantum Machine Intelligence}, vol.~3, no.~2,
  p.~23, 2021.

\bibitem{nath2021review}
R.~K. Nath, H.~Thapliyal, and T.~S. Humble, ``A review of machine learning
  classification using quantum annealing for real-world applications,''
  \emph{SN Computer science}, vol.~2, pp. 1--11, 2021.

\bibitem{hibat2021variational}
M.~Hibat-Allah, E.~M. Inack, R.~Wiersema, R.~G. Melko, and J.~Carrasquilla,
  ``Variational neural annealing,'' \emph{Nature Machine Intelligence}, vol.~3,
  no.~11, pp. 952--961, 2021.

\bibitem{warren2013adapting}
R.~H. Warren, ``Adapting the traveling salesman problem to an adiabatic quantum
  computer,'' \emph{Quantum information processing}, vol.~12, pp. 1781--1785,
  2013.

\bibitem{warren2020solving}
R.~H. Warren, ``Solving the traveling salesman problem on a quantum annealer,''
  \emph{SN Applied Sciences}, vol.~2, no.~1, p.~75, 2020.

\bibitem{SYRICHAS201752}
\BIBentryALTinterwordspacing
A.~Syrichas and A.~Crispin, ``Large-scale vehicle routing problems: Quantum
  annealing, tunings and results,'' \emph{Computers \& Operations Research},
  vol.~87, pp. 52--62, 2017. [Online]. Available:
  \url{https://www.sciencedirect.com/science/article/pii/S0305054817301260}
\BIBentrySTDinterwordspacing

\bibitem{inoue2021traffic}
D.~Inoue, A.~Okada, T.~Matsumori, K.~Aihara, and H.~Yoshida, ``Traffic signal
  optimization on a square lattice with quantum annealing,'' \emph{Scientific
  reports}, vol.~11, no.~1, pp. 1--12, 2021.

\bibitem{hussain2020optimal}
H.~Hussain, M.~B. Javaid, F.~S. Khan, A.~Dalal, and A.~Khalique, ``Optimal
  control of traffic signals using quantum annealing,'' \emph{Quantum
  Information Processing}, vol.~19, pp. 1--18, 2020.

\bibitem{denkena2021quantum}
B.~Denkena, F.~Schinkel, J.~Pirnay, and S.~Wilmsmeier, ``Quantum algorithms for
  process parallel flexible job shop scheduling,'' \emph{CIRP Journal of
  Manufacturing Science and Technology}, vol.~33, pp. 100--114, 2021.

\bibitem{stollenwerk2021agile}
T.~Stollenwerk, V.~Michaud, E.~Lobe, M.~Picard, A.~Basermann, and T.~Botter,
  ``Agile earth observation satellite scheduling with a quantum annealer,''
  \emph{IEEE Transactions on Aerospace and Electronic Systems}, vol.~57, no.~5,
  pp. 3520--3528, 2021.

\bibitem{ding2021implementation}
Y.~Ding, X.~Chen, L.~Lamata, E.~Solano, and M.~Sanz, ``Implementation of a
  hybrid classical-quantum annealing algorithm for logistic network design,''
  \emph{SN Computer Science}, vol.~2, pp. 1--9, 2021.

\bibitem{surveyclass}
M.~H. Alkawaz and A.~Silvarajoo, ``A survey on test case prioritization and
  optimization techniques in software regression testing,'' in \emph{2019 IEEE
  7th Conference on Systems, Process and Control (ICSPC)}, 2019, pp. 59--64.

\bibitem{tcs1}
T.~Nithya and S.~Chitra, ``Soft computing-based semi-automated test case
  selection using gradient-based techniques,'' \emph{Soft Computing}, vol.~24,
  no.~17, pp. 12\,981--12\,987, 2020.

\bibitem{tcs2}
D.~K. Yadav and S.~Dutta, ``Regression test case selection and prioritization
  for object oriented software,'' \emph{Microsystem Technologies}, vol.~26, pp.
  1463--1477, 2020.

\bibitem{tcs3}
\BIBentryALTinterwordspacing
T.~\c{C}\i{}ng\i{}l and H.~S\"{o}zer, ``Black-box test case selection by
  relating code changes with previously fixed defects,'' in \emph{Proceedings
  of the International Conference on Evaluation and Assessment in Software
  Engineering 2022}, ser. EASE '22.\hskip 1em plus 0.5em minus 0.4em\relax New
  York, NY, USA: Association for Computing Machinery, 2022, pp. 30--39.
  [Online]. Available: \url{https://doi.org/10.1145/3530019.3530023}
\BIBentrySTDinterwordspacing

\bibitem{zhao2020quantum}
\BIBentryALTinterwordspacing
J.~Zhao, ``Quantum software engineering: Landscapes and horizons,''
  \emph{CoRR}, vol. abs/2007.07047, 2020. [Online]. Available:
  \url{https://arxiv.org/abs/2007.07047}
\BIBentrySTDinterwordspacing

\bibitem{Serrano2022bookQSE}
\BIBentryALTinterwordspacing
M.~A. Serrano, R.~P{\'{e}}rez{-}Castillo, and M.~Piattini, Eds., \emph{Quantum
  Software Engineering}.\hskip 1em plus 0.5em minus 0.4em\relax Springer
  International Publishing, 2022. [Online]. Available:
  \url{https://doi.org/10.1007/978-3-031-05324-5}
\BIBentrySTDinterwordspacing

\bibitem{MiranskyyICSE2020}
\BIBentryALTinterwordspacing
A.~Miranskyy, L.~Zhang, and J.~Doliskani, ``Is your quantum program bug-free?''
  in \emph{Proceedings of the ACM/IEEE 42nd International Conference on
  Software Engineering: New Ideas and Emerging Results}, ser. ICSE-NIER
  '20.\hskip 1em plus 0.5em minus 0.4em\relax New York, NY, USA: Association
  for Computing Machinery, 2020, pp. 29--32. [Online]. Available:
  \url{https://doi.org/10.1145/3377816.3381731}
\BIBentrySTDinterwordspacing

\bibitem{MiranskyyICSE19}
\BIBentryALTinterwordspacing
A.~Miranskyy and L.~Zhang, ``On testing quantum programs,'' in
  \emph{Proceedings of the 41st International Conference on Software
  Engineering: New Ideas and Emerging Results}, ser. ICSE-NIER '19.\hskip 1em
  plus 0.5em minus 0.4em\relax IEEE Press, 2019, pp. 57--60. [Online].
  Available: \url{https://doi.org/10.1109/ICSE-NIER.2019.00023}
\BIBentrySTDinterwordspacing

\bibitem{Garcia2023}
\BIBentryALTinterwordspacing
A.~Garc{\'i}a de~la Barrera, I.~Garc{\'i}a-Rodríguez~de Guzm{\'a}n, M.~Polo,
  and M.~Piattini, ``Quantum software testing: State of the art,''
  \emph{Journal of Software: Evolution and Process}, vol.~35, no.~4, p. e2419,
  2023. [Online]. Available:
  \url{https://onlinelibrary.wiley.com/doi/abs/10.1002/smr.2419}
\BIBentrySTDinterwordspacing

\bibitem{search}
\BIBentryALTinterwordspacing
X.~Wang, P.~Arcaini, T.~Yue, and S.~Ali, ``{QuSBT}: Search-based testing of
  quantum programs,'' in \emph{Proceedings of the ACM/IEEE 44th International
  Conference on Software Engineering: Companion Proceedings}, ser. ICSE
  '22.\hskip 1em plus 0.5em minus 0.4em\relax New York, NY, USA: Association
  for Computing Machinery, 2022, pp. 173--177. [Online]. Available:
  \url{https://doi.org/10.1145/3510454.3516839}
\BIBentrySTDinterwordspacing

\bibitem{mutation2}
\BIBentryALTinterwordspacing
X.~Wang, T.~Yu, P.~Arcaini, T.~Yue, and S.~Ali, ``Mutation-based test
  generation for quantum programs with multi-objective search,'' in
  \emph{Proceedings of the Genetic and Evolutionary Computation Conference},
  ser. GECCO '22.\hskip 1em plus 0.5em minus 0.4em\relax New York, NY, USA:
  Association for Computing Machinery, 2022, pp. 1345--1353. [Online].
  Available: \url{https://doi.org/10.1145/3512290.3528869}
\BIBentrySTDinterwordspacing

\bibitem{fuzz}
J.~Wang, F.~Ma, and Y.~Jiang, ``Poster: Fuzz testing of quantum program,'' in
  \emph{2021 14th IEEE Conference on Software Testing, Verification and
  Validation (ICST)}, 2021, pp. 466--469.

\bibitem{mutation}
E.~Mendiluze, S.~Ali, P.~Arcaini, and T.~Yue, ``Muskit: A mutation analysis
  tool for quantum software testing,'' in \emph{2021 36th IEEE/ACM
  International Conference on Automated Software Engineering (ASE)}, 2021, pp.
  1266--1270.

\bibitem{mutation1}
\BIBentryALTinterwordspacing
D.~Fortunato, J.~Campos, and R.~Abreu, ``Mutation testing of quantum programs
  written in {QISKit},'' in \emph{Proceedings of the ACM/IEEE 44th
  International Conference on Software Engineering: Companion Proceedings},
  ser. ICSE '22.\hskip 1em plus 0.5em minus 0.4em\relax New York, NY, USA:
  Association for Computing Machinery, 2022, pp. 358--359. [Online]. Available:
  \url{https://doi.org/10.1145/3510454.3528649}
\BIBentrySTDinterwordspacing

\bibitem{Rui_mutation}
\BIBentryALTinterwordspacing
D.~Fortunato, J.~Campos, and R.~Abreu, ``{QMutPy}: A mutation testing tool for quantum algorithms and
  applications in {Qiskit},'' in \emph{Proceedings of the 31st ACM SIGSOFT
  International Symposium on Software Testing and Analysis}, ser. ISSTA
  2022.\hskip 1em plus 0.5em minus 0.4em\relax New York, NY, USA: Association
  for Computing Machinery, 2022, pp. 797--800. [Online]. Available:
  \url{https://doi.org/10.1145/3533767.3543296}
\BIBentrySTDinterwordspacing

\bibitem{Fortunato2022}
D.~Fortunato, J.~Campos, and R.~Abreu, ``Mutation testing of quantum programs: A case study with {Qiskit},''
  \emph{IEEE Transactions on Quantum Engineering}, vol.~3, pp. 1--17, 2022.

\bibitem{Rui_metamorphic}
\BIBentryALTinterwordspacing
R.~Abreu, J.~a.~P. Fernandes, L.~Llana, and G.~Tavares, ``Metamorphic testing
  of oracle quantum programs,'' in \emph{Proceedings of the 3rd International
  Workshop on Quantum Software Engineering}, ser. Q-SE '22.\hskip 1em plus
  0.5em minus 0.4em\relax New York, NY, USA: Association for Computing
  Machinery, 2023, pp. 16--23. [Online]. Available:
  \url{https://doi.org/10.1145/3528230.3529189}
\BIBentrySTDinterwordspacing

\bibitem{morphq}
\BIBentryALTinterwordspacing
M.~Paltenghi and M.~Pradel, ``{MorphQ}: Metamorphic testing of quantum
  computing platforms,'' \emph{CoRR}, vol. abs/2206.01111, 2022. [Online].
  Available: \url{https://doi.org/10.48550/arXiv.2206.01111}
\BIBentrySTDinterwordspacing

\bibitem{Combinatorial}
X.~Wang, P.~Arcaini, T.~Yue, and S.~Ali, ``Application of combinatorial testing
  to quantum programs,'' in \emph{2021 IEEE 21st International Conference on
  Software Quality, Reliability and Security (QRS)}, 2021, pp. 179--188.

\bibitem{spieker2017reinforcement}
\BIBentryALTinterwordspacing
H.~Spieker, A.~Gotlieb, D.~Marijan, and M.~Mossige, ``Reinforcement learning
  for automatic test case prioritization and selection in continuous
  integration,'' in \emph{Proceedings of the 26th ACM SIGSOFT International
  Symposium on Software Testing and Analysis}, ser. ISSTA 2017.\hskip 1em plus
  0.5em minus 0.4em\relax New York, NY, USA: Association for Computing
  Machinery, 2017, pp. 12--22. [Online]. Available:
  \url{https://doi.org/10.1145/3092703.3092709}
\BIBentrySTDinterwordspacing

\bibitem{energyimpact}
``D-wave hybrid,'' D-wave systems, Tech. Rep., 2023.

\bibitem{dwave}
\BIBentryALTinterwordspacing
M.~Catherine and F.~Pau, ``The d-wave advantage system: An overview,'' D-Wave,
  Tech. Rep., 2022. [Online]. Available:
  \url{https://www.dwavesys.com/media/s3qbjp3s/14-1049a-a_the_d-wave_advantage_system_an_overview.pdf}
\BIBentrySTDinterwordspacing

\bibitem{arcuri2011practical}
\BIBentryALTinterwordspacing
A.~Arcuri and L.~Briand, ``A practical guide for using statistical tests to
  assess randomized algorithms in software engineering,'' in \emph{Proceedings
  of the 33rd International Conference on Software Engineering}, ser. ICSE
  '11.\hskip 1em plus 0.5em minus 0.4em\relax New York, NY, USA: Association
  for Computing Machinery, 2011, pp. 1--10. [Online]. Available:
  \url{https://doi.org/10.1145/1985793.1985795}
\BIBentrySTDinterwordspacing

\bibitem{abbott2019hybrid}
A.~A. Abbott, C.~S. Calude, M.~J. Dinneen, and R.~Hua, ``A hybrid
  quantum-classical paradigm to mitigate embedding costs in quantum
  annealing,'' \emph{International Journal of Quantum Information}, vol.~17,
  no.~05, p. 1950042, 2019.

\bibitem{Calaza2021}
\BIBentryALTinterwordspacing
C.~D. Gonzalez~Calaza, D.~Willsch, and K.~Michielsen, ``Garden optimization
  problems for benchmarking quantum annealers,'' \emph{Quantum Information
  Processing}, vol.~20, no.~9, sep 2021. [Online]. Available:
  \url{https://doi.org/10.1007/s11128-021-03226-6}
\BIBentrySTDinterwordspacing


\bibitem{energygap}
A.~Rajak, S.~Suzuki, A.~Dutta, and B.~K. Chakrabarti, ``Quantum annealing: an
  overview,'' \emph{Philosophical Transactions of the Royal Society A}, vol.
  381, no. 2241, p. 20210417, 2023.

\bibitem{choi2008minor}
V.~Choi, ``Minor-embedding in adiabatic quantum computation: I. the parameter
  setting problem,'' \emph{Quantum Information Processing}, vol.~7, pp.
  193--209, 2008.

\bibitem{emb1}
S.~Zbinden, A.~B{\"a}rtschi, H.~Djidjev, and S.~Eidenbenz, ``Embedding
  algorithms for quantum annealers with chimera and pegasus connection
  topologies,'' in \emph{High Performance Computing}, P.~Sadayappan, B.~L.
  Chamberlain, G.~Juckeland, and H.~Ltaief, Eds.\hskip 1em plus 0.5em minus
  0.4em\relax Cham: Springer International Publishing, 2020, pp. 187--206.

\bibitem{emb2}
P.~Thai, M.~T. Thai, T.~Vu, and T.~N. Dinh, ``Fasthare: Fast hamiltonian
  reduction for large-scale quantum annealing,'' in \emph{2022 IEEE
  International Conference on Quantum Computing and Engineering (QCE)}, 2022,
  pp. 114--124.

\bibitem{emb3}
K.~Boothby, P.~Bunyk, J.~Raymond, and A.~Roy, ``Next-generation topology of
  d-wave quantum processors,'' 2020.

\bibitem{emb4}
A.~Barbosa, E.~Pelofske, G.~Hahn, and H.~N. Djidjev, ``Optimizing
  embedding-related quantum annealing parameters for reducing hardware bias,''
  in \emph{Parallel Architectures, Algorithms and Programming}, L.~Ning,
  V.~Chau, and F.~Lau, Eds.\hskip 1em plus 0.5em minus 0.4em\relax Singapore:
  Springer Singapore, 2021, pp. 162--173.

\bibitem{parallel}
E.~Pelofske, G.~Hahn, and H.~N. Djidjev, ``Parallel quantum annealing,''
  \emph{Scientific Reports}, vol.~12, no.~1, pp. 1--11, 2022.

\bibitem{Wohlin2012}
C.~Wohlin, P.~Runeson, M.~Hst, M.~C. Ohlsson, B.~Regnell, and A.~Wessln,
  \emph{Experimentation in Software Engineering}.\hskip 1em plus 0.5em minus
  0.4em\relax Springer Publishing Company, Incorporated, 2012.

\end{thebibliography}

\end{document}